\documentclass[pra,reprint, longbibliography]{revtex4-2}
\usepackage{braket}
\usepackage{graphicx}
\usepackage{amsmath}
\usepackage{hyperref}
\usepackage[all]{hypcap}
\hypersetup{colorlinks=True}
\usepackage{color}
\usepackage{subfigure}
\usepackage{placeins}
\usepackage{times}
\usepackage{bbold}
\DeclareMathOperator{\Tr}{Tr}

\begin{document}
\author{Xin H. H. Zhang}
\email{xin.z@duke.edu}
\affiliation{Department of Physics, Duke University, P.O.\,Box 90305, Durham, NC 27708-0305, USA}
\author{Harold U. Baranger}
\affiliation{Department of Physics, Duke University, P.O.\,Box 90305, Durham, NC 27708-0305, USA}
\date{March 24, 2021}

\title{Driven-dissipative phase transition in a Kerr oscillator: \\From semiclassical $\mathcal{PT}$ symmetry to quantum fluctuations}
\begin{abstract}
%
%
%
We study a minimal model that has a driven-dissipative quantum phase transition, namely a Kerr non-linear oscillator subject to driving and dissipation. Using mean-field theory, exact diagonalization, and the Keldysh formalism, we analyze the critical phenomena in this system, showing which aspects can be captured by each approach and how the approaches complement each other. Then critical scaling and finite-size scaling are calculated analytically using the quantum Langevin equation. The physics contained in this simple model is surprisingly rich: it includes a continuous phase transition, $Z_{2}$ symmetry breaking, $\mathcal{PT}$ symmetry, state squeezing, and critical fluctuations. Due to its simplicity and solvability, this model can serve as a paradigm for exploration of open quantum many-body physics.
\end{abstract}

\maketitle

\section{Introduction}

Quantum phase transitions (QPT) have been one of the central themes of many-body physics \cite{SachdevBook2011,CarrBook}. In closed systems governed by a Hamiltonian, a QPT signifies an abrupt qualitative change of the ground state wavefunction. Recently, due to interest in non-equilibrium quantum many-body physics \cite{EisertNatPhys2015}, phase transitions in open systems have attracted increasing attention. Because of coupling with an external environment, an open system is governed by non-unitary dynamics. Nevertheless, there can be an analogous abrupt change of a system's steady state: this is a  phase transition in an open quantum system (e.g.\,\cite{DrummondJPA1980,CarmichaelJPB1980,WernerPRL2005,CapriottiPRL2005,MitraPRL2006,ProsenPRL2008,DiehlPRL2010,KesslerPRA2012,TorrePRA2013,CarmichaelPRX2015,FinkPRX2017,FitzpatrickPRX2017,RodriguezPRL2017,FinkNatPhys2018,BrookesArxiv2019}). However, because a general framework for non-equilibrium physics is lacking, QPT in open quantum systems are much less well understood. To make the principal mechanism more transparent, here we study a simple and solvable system: an oscillator with Kerr non-linear interaction, two-photon driving, and single-photon loss. The competition between driving and dissipation leads to a second-order phase transition in the steady state. The system illustrates the basic principles of a driven-dissipative phase transition and contains rich physics. Since an analytical solution is given, it can serve as a paradigmatic model for the study of open quantum many-body physics.

For a Kerr oscillator subject to single-photon driving, there exists a first-order phase transition and hysteresis, which has been studied theoretically using the generalized P-representation \cite{DrummondJPA1980,BartoloPRA2016}, the quantum-absorber method \cite{RobertsPRX2020}, quantum activation \cite{DykmanBook2012}, and numerical diagonalization \cite{CasteelsPRA2016,CasteelsPRA2017,MingantiPRA2018,KrimerPRL2019,HeugelPRL2019}. Experimentally, it has been realized using a semiconductor optical cavity \cite{RodriguezPRL2017,FinkNatPhys2018,GengPRL2020} and in circuit QED \cite{BrookesArxiv2019}. When two-photon driving is used, the system has a parity symmetry; it has been shown theoretically that this symmetry can be spontaneously broken in the steady state, leading to a continuous phase transition \cite{BartoloPRA2016,MingantiPRA2018}. Recently, this symmetry breaking has found applications in quantum error correction \cite{LieuPRL2020}.  Furthermore, phase transitions in either a few or a lattice of coupled Kerr oscillators have also been explored \cite{BoitePRL2013,FossFeig2017,MascarenhasArxiv2017,CasteelsPRA2017b,RotaPRL2019,VerstraelenPRR2020,VerstraelenPRA2020}. Dissipative phase transitions have also been studied in a dissipative cavity with other types of non-linearity, such as coupling with atoms (e.g.\ \cite{TorrePRA2013,CarmichaelPRX2015,FinkPRX2017,VukicsQuantum2019,HwangPRA2018,ZhuPRL2020,ParaoanuPRA2020}).   

Previous approaches to the phase transition in a Kerr oscillator are usually limited to finite system size. Here, using the Keldysh formalism, we access the thermodynamic limit directly. At both the semi-classical and quantum levels, we reveal the mechanism behind the driven-dissipative second-order phase transition, which is crucial for understanding general cases. At the semi-classical level, we find that the phase transition is connected to an emergent $\mathcal{PT}$ symmetry of its dynamics (sections \ref{sec:MFT} and \ref{sec:PT}). Then to show the spectrum of the Lindbladian, symmetry breaking in the steady state manifold, and the validity of the semi-classical approach, quantum solutions are provided numerically (section \ref{sec:SpectrumL}). In the thermodynamic limit (i.e.\ the weak non-linearity limit), using the Keldysh formalism, we provide analytical solutions for the spectral function and power spectrum, which are experimentally accessible and provide spectral signatures of the criticality (sections \ref{sec:Keldysh}, \ref{sec:SpectralFunction}, and \ref{sec:PowerSpectrum}). Finally, critical and finite-size scaling exponents are calculated analytically using a quantum Langevin equation (section \ref{sec:QLangevin}).













\begin{figure}[bt]
	\center
	\includegraphics[width=0.3\textwidth]{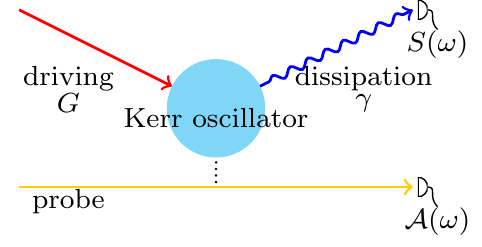}
	\put(-10,30){(a)}
	
	\includegraphics[width=0.35\textwidth]{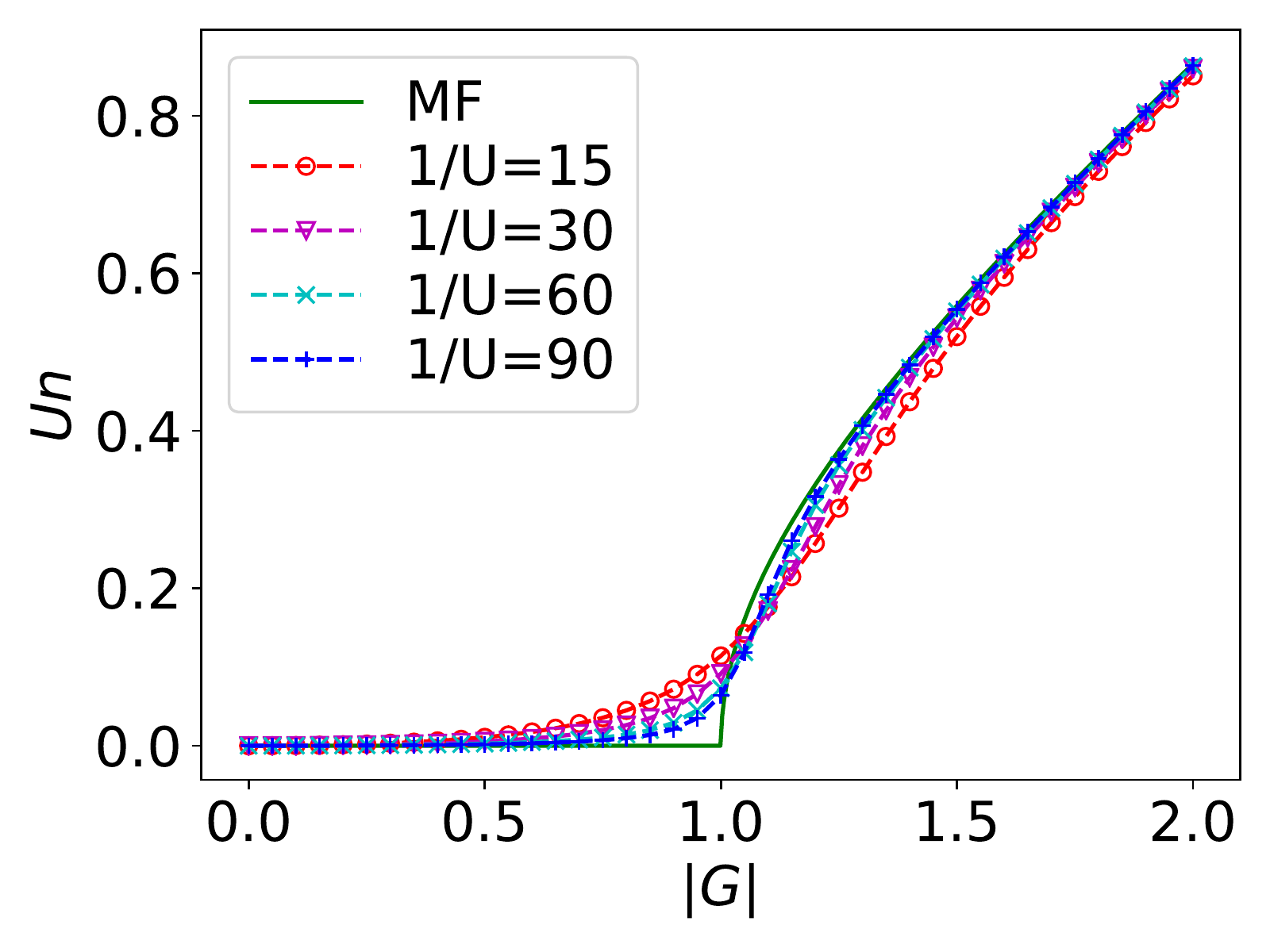}
    \put(-25,35){(b)}
	\caption{(a) Schematic of the setup. Detection of the dissipative photons gives their power spectrum $S(\omega)$ and detection of the scattered probing light gives the absorption spectrum $\mathcal{A}(\omega)$. (b) Order parameter $Un$ v.s.\,driving strength $|G|$ for MF theory and for different values of $U$ obtained with diagonalization. 
	(Values of $1/U$ in the key refer to curves from top to bottom for $|G|\!<\!1$.)
	}
	\label{OrderParameter}
\end{figure}

\section{The Model}

The system we study is a non-linear oscillator subject to two-photon (parametric) driving and single-photon loss [Fig.\,\ref{OrderParameter}(a)]. The unitary dynamics can be described by the Hamiltonian (in a rotating frame) 
\begin{equation}
H = -\omega_{d} a^{\dagger} a + \frac{U}{2} a^{\dagger} a^{\dagger} a a + \left(\frac{G}{4} a^{\dagger}a^{\dagger} + \text{h.c.}\right),
\end{equation}
where the detuning $\omega_{d} = \omega_{G}/2 - \omega_{c}$ with $\omega_{G}$ and $\omega_{c}$ being the frequency of the driving and the cavity, respectively  \footnote{The driving strength $G/4$ is defined to make the driving strength comparable to the loss rate of single photons as shown later in \eqref{aavr}.}. 
With dissipation the system is then described by a Lindblad master equation
\begin{equation} \label{Lindblad}
\frac{d}{dt} \rho = -i [H, \rho] + \gamma D(a) [\rho],
\end{equation}
where $D(a) [\rho] = a \rho a^{\dagger} - \frac{1}{2} \{ \rho, a^{\dagger} a \}$. For abbreviation, the right-hand side (RHS) of \eqref{Lindblad} is denoted as $\mathcal{L} \rho$, where $\mathcal{L}$ is a superoperator called the Lindbladian. It can be seen that $\mathcal{L}$ is invariant under a $Z_{2}$ transformation $a^{(\dagger)} \rightarrow  -a^{(\dagger)}$. This model could be realized using several experimental platforms such as semiconductor microcavities (e.g.\ \cite{RodriguezPRL2017,FinkNatPhys2018,GengPRL2020}) or quantum circuits (e.g.\ \cite{LeghtasScience2015,BrookesArxiv2019,BlaisCircuitReview2020}).

\section{Mean Field Theory}\label{sec:MFT}

From \eqref{Lindblad}, the equation of motion for the expectation value of $a$ is
\begin{equation} \label{aavr}
\frac{d}{dt} \braket{a} =  i \omega_{d}  \braket{a} -i U \braket{a^{\dagger} a a} - i \frac{G}{2} \braket{a^{\dagger}} - \frac{\gamma}{2} \braket{a}.
\end{equation} 
By making the approximation that $\braket{a^{\dagger} a a} \approx \braket{a^{\dagger}}\braket{a}\braket{a}$ (see e.g.\,\cite{DrummondJPA1980,BartoloPRA2016}), we obtain the semi-classical mean field (MF) equation of motion for $\alpha \equiv \braket{a}$,
\begin{equation}\label{MFT}
\frac{d}{dt} \alpha =  \big( i \omega_{d}  -i U |\alpha|^2 - \frac{\gamma}{2} \big) \alpha - i \frac{G}{2} \alpha^{*}.
\end{equation}
The steady state is obtained by setting RHS $=0$,
which leads to the solution for the occupation number $n = |\alpha|^2$ (assuming $\omega_{d} \geq 0$): when $|G|<\gamma$, $n=0$; when $\gamma \leq |G|\leq \sqrt{\gamma^2+4 \omega_{d} ^2}$, $n=0$ or $(\omega_{d} +\frac{1}{2}\sqrt{|G|^2-\gamma^2})/U$; when $|G|>\sqrt{\gamma^2+4 \omega_{d} ^2}$, $n = (\omega_{d} +\frac{1}{2}\sqrt{|G|^2-\gamma^2})/U$. 


When the driving is resonant ($\omega_{d} =0$), which is the focus of this paper, we see that MF theory predicts a second-order phase transition at the critical driving strength $|G_{c}| = \gamma$. We define the order parameter as $\phi = Un$. Then
\begin{equation} \label{MFTphi}
\phi = \begin{cases} 
0 & |G|<\gamma, \\[0.1cm]
 \frac{1}{2}\sqrt{|G|^2-\gamma^2} & |G|>\gamma.
\end{cases}
\end{equation}
By substituting the solution for $\phi$ back into \eqref{MFT}, one finds that the $Z_{2}$ symmetry is broken when $G>\gamma$ since the steady state amplitude is  $\alpha_{s} = \pm \sqrt{n} e^{i\theta}$, where the phase factor is given by 
\begin{equation}\label{phasetheta}
e^{i2\theta} = -(\sqrt{|G|^2-\gamma^2}+i\gamma)/G^{*}.    
\end{equation} In calculations, we take $\gamma=1$ unless specified otherwise.

\begin{figure}[b]
	\center
	\includegraphics[width=0.23\textwidth]{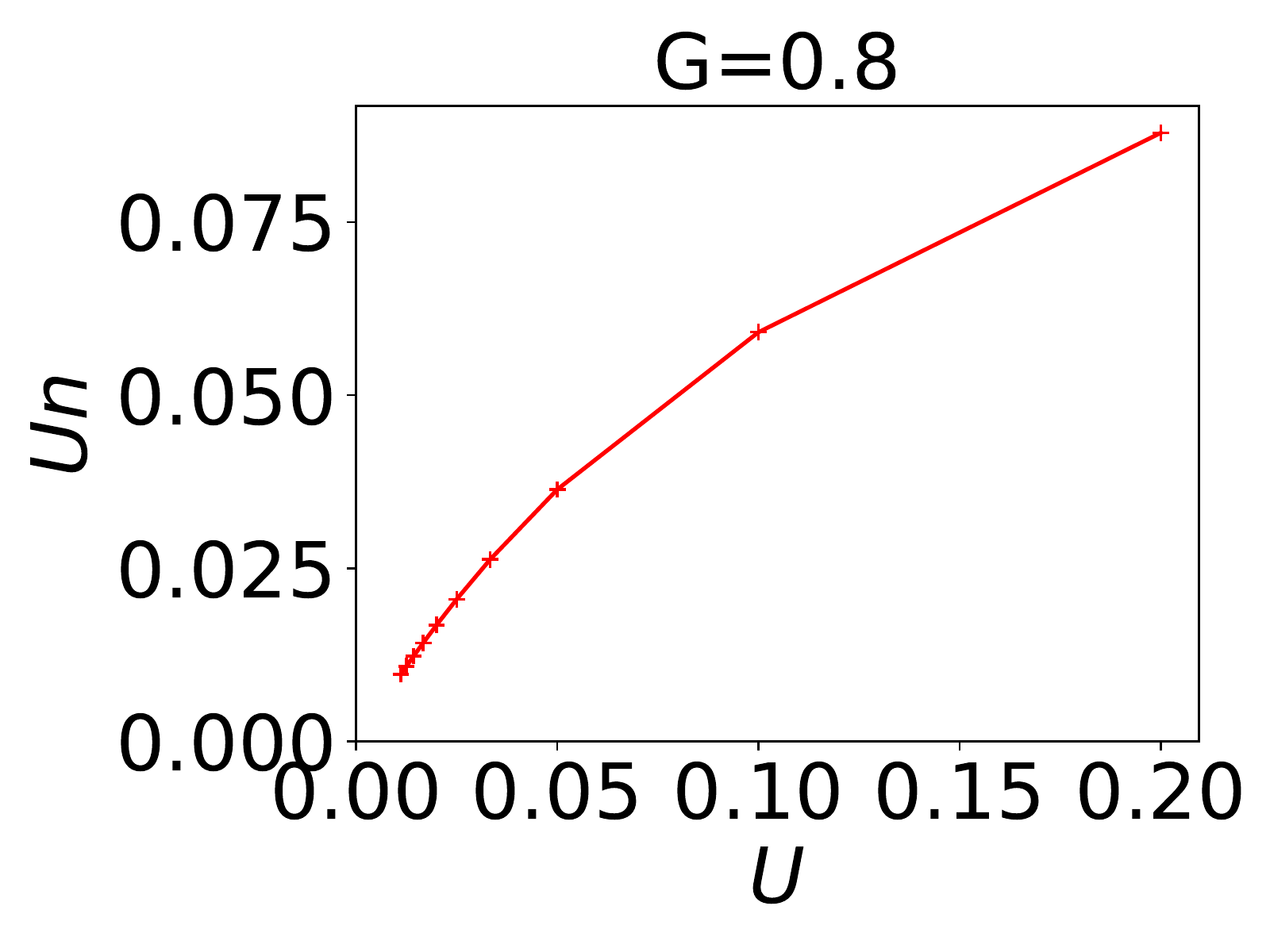}
		\includegraphics[width=0.23\textwidth]{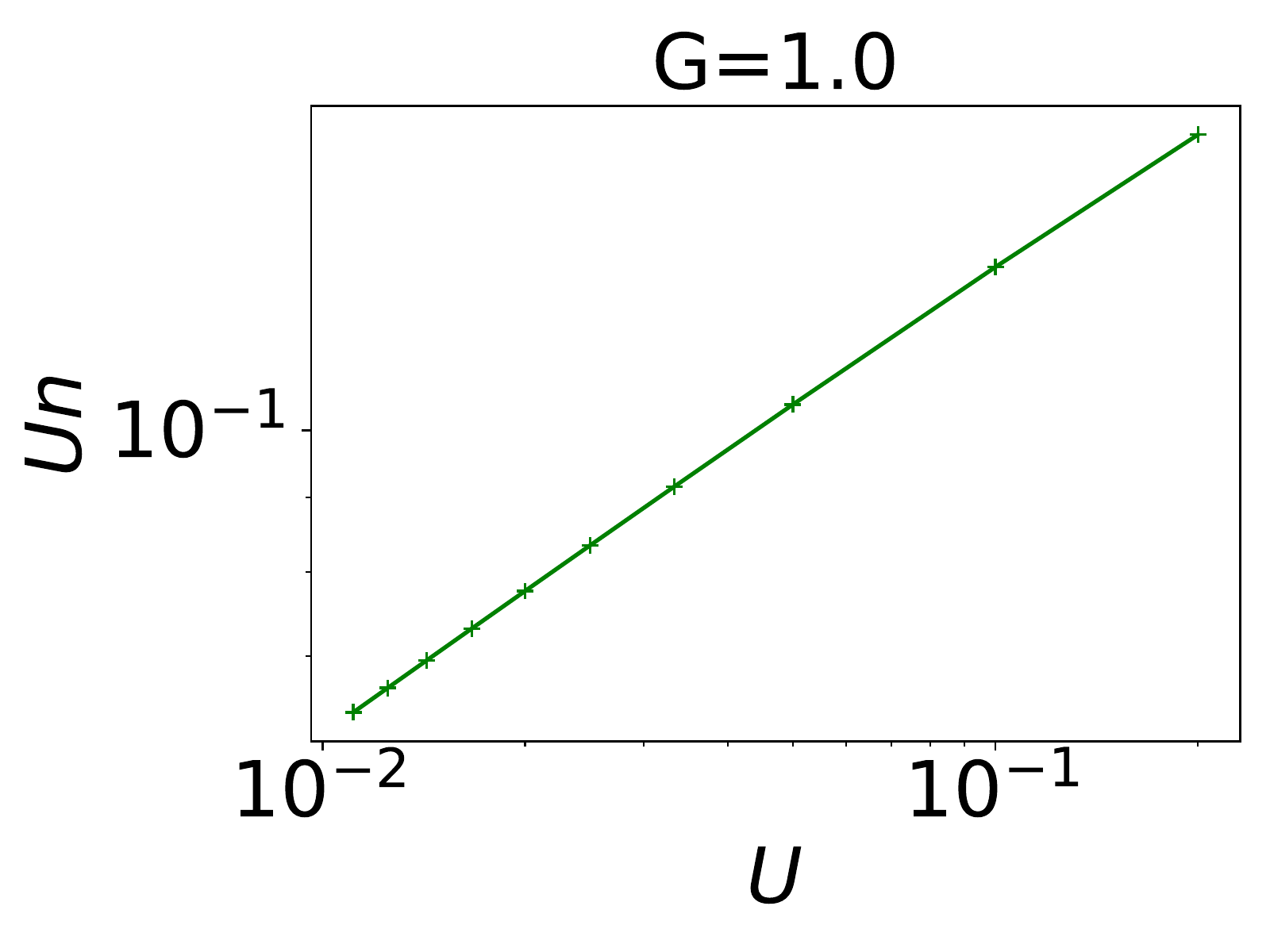}
	\put(-143,25){(a)}
	\put(-23,25){(b)}
	
	\includegraphics[width=0.23\textwidth]{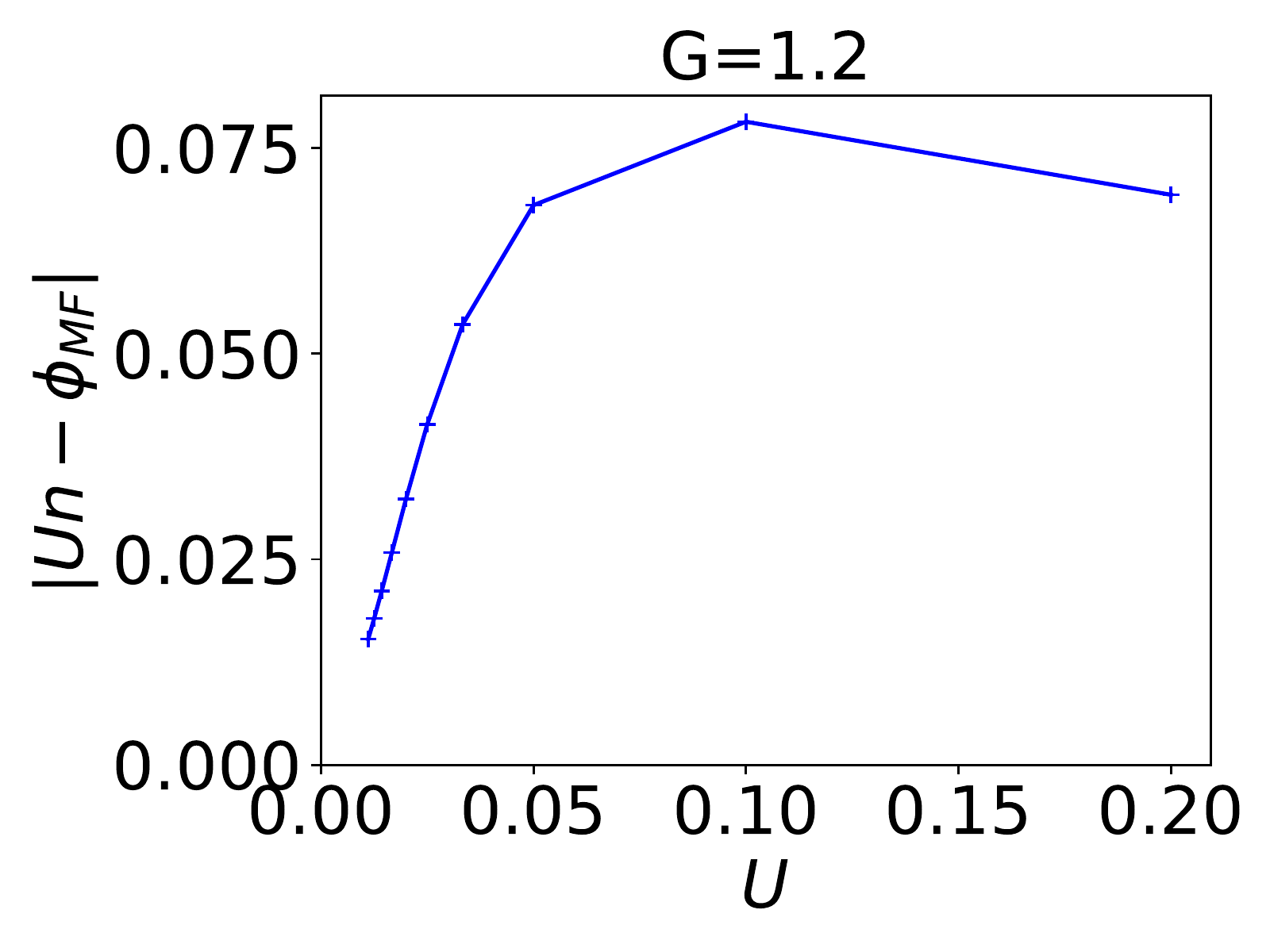}
	\includegraphics[width=0.23\textwidth]{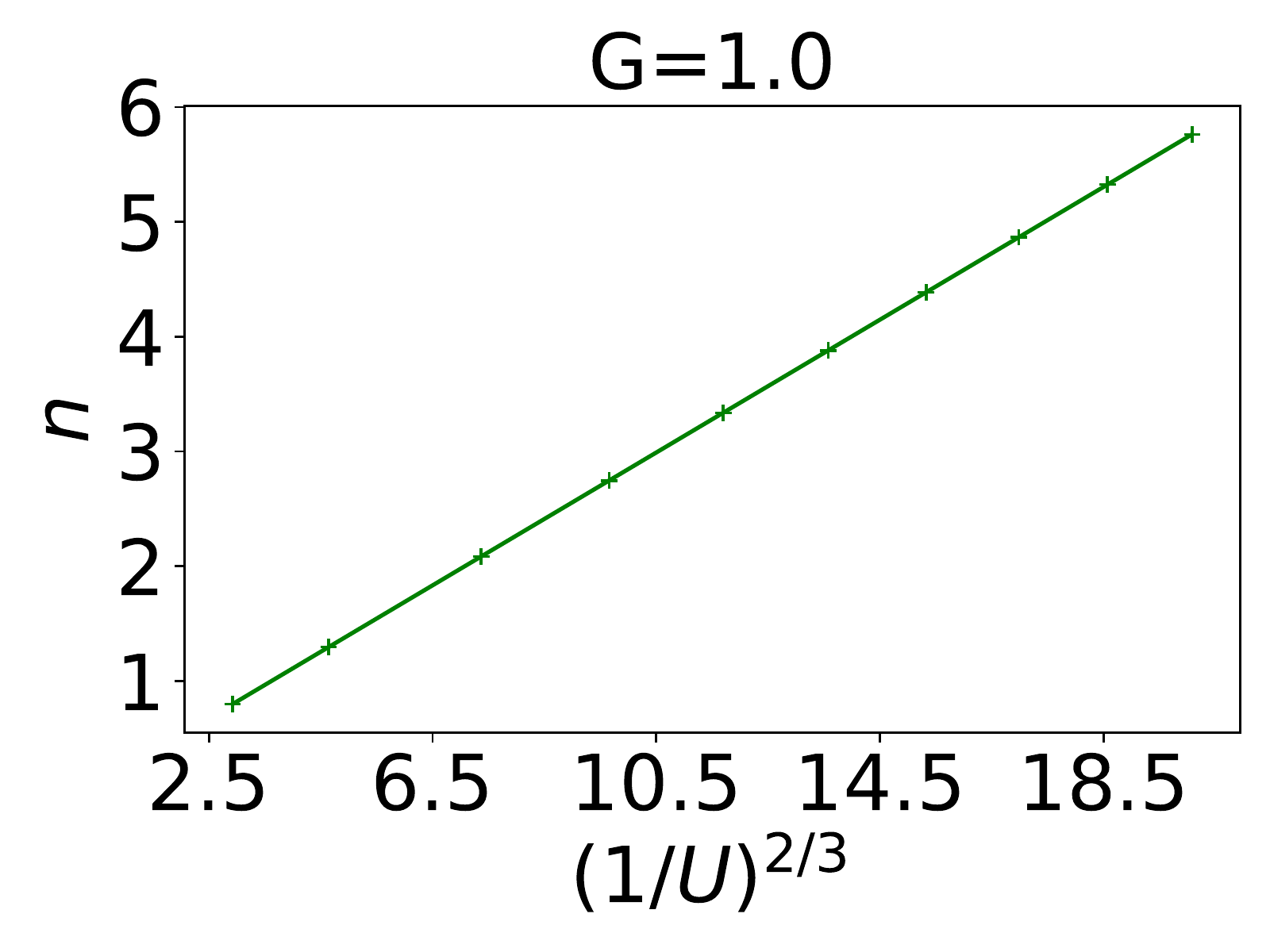}
	\put(-140,25){(c)}
	\put(-20,25){(d)}
	\caption{$|\phi_{\text{ED}} - \phi_{\text{MF}}|$ vs.\,$U$ for (a) $G=0.8$, (b) $G=1.0$ (log-log scale), and (c) $G=1.2$. (b) shows clearly a power law scaling at critical point. (d) The linearity between $n$ and $U^{-2/3}$, which means $Un \propto U^{1/3}$.  
	}
	\label{FiniteSizeOrderParameter}
\end{figure}

The number of photons in the cavity $n$ is of order $\gamma/U$, which means that the thermodynamic limit is the limit of infinitesimal interaction $U/\gamma \rightarrow 0^{+}$. Note that even though the interaction strength $U$ is small, the interaction term $\frac{U}{2}a^{\dagger}a^{\dagger}aa$ is of the order $Un^2 = \phi n$, which is comparable to other terms in the Lindbladian. To check the validity of MF theory, we diagonalize the Lindbladian $\mathcal{L}$ exactly (see Appendix \ref{Appdix:ED}). As shown in Fig.\ref{OrderParameter}(b), as $U$ approaches $0$, the value $U \braket{a^{\dagger}a}$ approaches the MF result. Fig.\,\ref{FiniteSizeOrderParameter} shows  $|\phi_{\text{ED}} - \phi_{\text{MF}}|$ as a function of $U$,  which indeed vanishes as $U\rightarrow 0^{+}$. At the critical point, it scales as a power law $Un \propto U^{1/3}$, which is calculated analytically later in section \ref{sec:QLangevin}. 

\section{Parity-Time ($\mathcal{PT}$) symmetry}\label{sec:PT}

At the MF level, this phase transition is related to the underlying structure of the MF equation of motion, which has a $\mathcal{PT}$ symmetry \cite{BenderRPP2007,ElGanainyNatPhys2018}. Eq.\,\eqref{MFT} can be written as
\begin{equation}\label{PTEOM}
\frac{d}{dt} \begin{pmatrix}
\alpha    \\
\alpha^{*}
\end{pmatrix}
= M
\begin{pmatrix}
\alpha    \\
\alpha^{*}
\end{pmatrix},
\end{equation}
where
\begin{equation}\label{M}
M =
\begin{pmatrix}
-i \phi - \frac{\gamma}{2} & -i G   \\
iG^{*} & i \phi - \frac{\gamma}{2}
\end{pmatrix}.
\end{equation}
$M$ is $\mathcal{PT}$-symmetric \footnote{The name ``parity'' of $\mathcal{PT}$ follows the convention in Ref.\,\cite{BenderRPP2007}, which is different from the parity of the Lindbladian mentioned above.} since it is invariant under the exchange of modes combined with complex conjugation (e.g.\ \cite{BenderRPP2007,ElGanainyNatPhys2018,WangPRA2019}). Here this symmetry emerges in the thermodynamic limit, when fluctuations are negligible and the MF equation of motion becomes exact. 

\begin{figure}[tb]
	\center
	\includegraphics[width=0.5\textwidth]{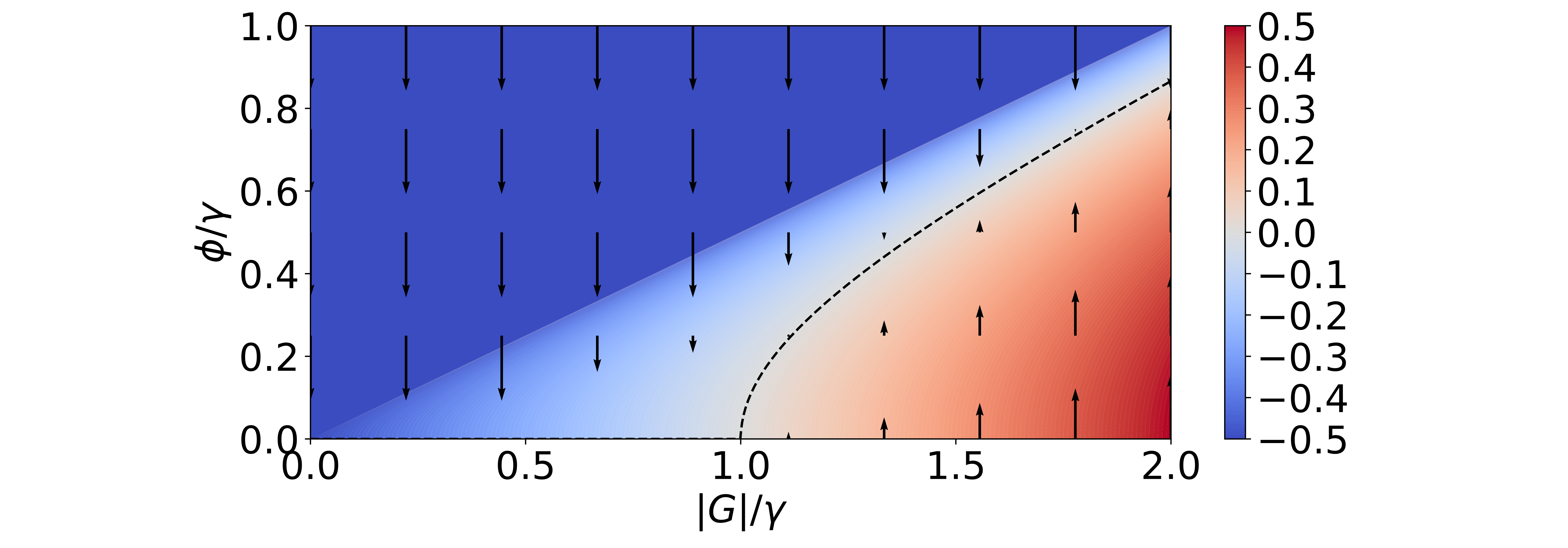}
	\caption{$\Re \Gamma_{+}$ as a function of $\phi$ and $|G|$. The arrows are proportional to the magnitude of $\Re \Gamma_{+}$ and they point up (down) for positive (negative) sign. The dashed line marks the stabilized solution where $\Re \Gamma_{+}=0$. The boundary between lower right ($\Re \Gamma_{+}> - \gamma/2$) and upper left ($\Re \Gamma_{+}= - \gamma/2$) is a line of exceptional points.}
	\label{PTflow}
\end{figure}

This emergent $\mathcal{PT}$-symmetry guarantees that eigenvalues of $M$ have the form  $\Gamma_{\pm} =  - \gamma/2 \pm \xi$ with $\xi = \sqrt{|G|^2-4\phi^2}/2$. $\xi$ is purely real when $|G|>2 \phi$ ($\mathcal{PT}$ symmetry unbroken) while purely imaginary when  $|G|<2 \phi$ ($\mathcal{PT}$ symmetry broken). $|G|=2\phi$ defines a line of exceptional points where $\xi=0$ and two eigenmodes coalesce. Commonly studied $\mathcal{PT}$-symmetric systems have $\gamma=0$ in which the exceptional points are therefore at the boundary between exponentially increasing/decreasing modes and oscillatory modes \cite{BenderRPP2007,ElGanainyNatPhys2018}. Since here $\gamma\neq 0$, $\Re \Gamma_{+}$ in contrast determines the dynamics at long times.  $\Re \Gamma_{+}<(>)\,0$ leads to exponential decrease (increase) of $\alpha$ and $\phi$. Then as shown in Fig.\,\ref{PTflow}, this $\mathcal{PT}$ symmetry of $M$ leads to a self-stabilizing mechanism: the dynamics drives the solution to the line  $\Re \Gamma_{+}=0$, which is the same as the MF solution.

Note that here we consider the $\mathcal{PT}$ symmetry at the MF level; $\mathcal{PT}$ symmetry of Lindbladian, a separate issue, has also been found to be connected with dissipative phase transitions \cite{ProsenPRL2012,HuybrechtsPRB2020,HuberPRA2020,HuberSciPostPhys2020,CurtisArxiv2020}. 

\section{Spectrum of Lindbladian}\label{sec:SpectrumL}

To understand the physics of this system at the quantum level and see how many-body effects emerge in the thermodynamic limit, we study the spectrum of $\mathcal{L}$ using exact diagonalization (ED):
\begin{equation}
\mathcal{L} \sigma_{i} = \lambda_{i} \sigma_{i},
\end{equation} 
where the eigenvalues $\{ \lambda_{i} | i=0,\ldots,n_{\text{max}} \}$ have been sorted in descending order of their real parts and $n_{\text{max}}$ is the occupation number cutoff. For a Lindbladian, the real parts of the eigenvalues are non-positive ($\Re \lambda_{i} \leq 0$), and there must exist at least one $\lambda_{i}=0$, which correspond to the steady states (e.g.\,\cite{KesslerPRA2012, MingantiPRA2018}). The gap $\Delta$ between the second largest and $0$ gives a smallest decay rate towards the steady state. This rate therefore dominates the long-time dynamics. As in the closing of an energy gap in a Hamiltonian system, the closing of this gap signals degeneracy of the steady state manifold. Such degeneracy happens at a quantum phase transition.

\begin{figure}[t]
	\center
	\includegraphics[width=0.23\textwidth]{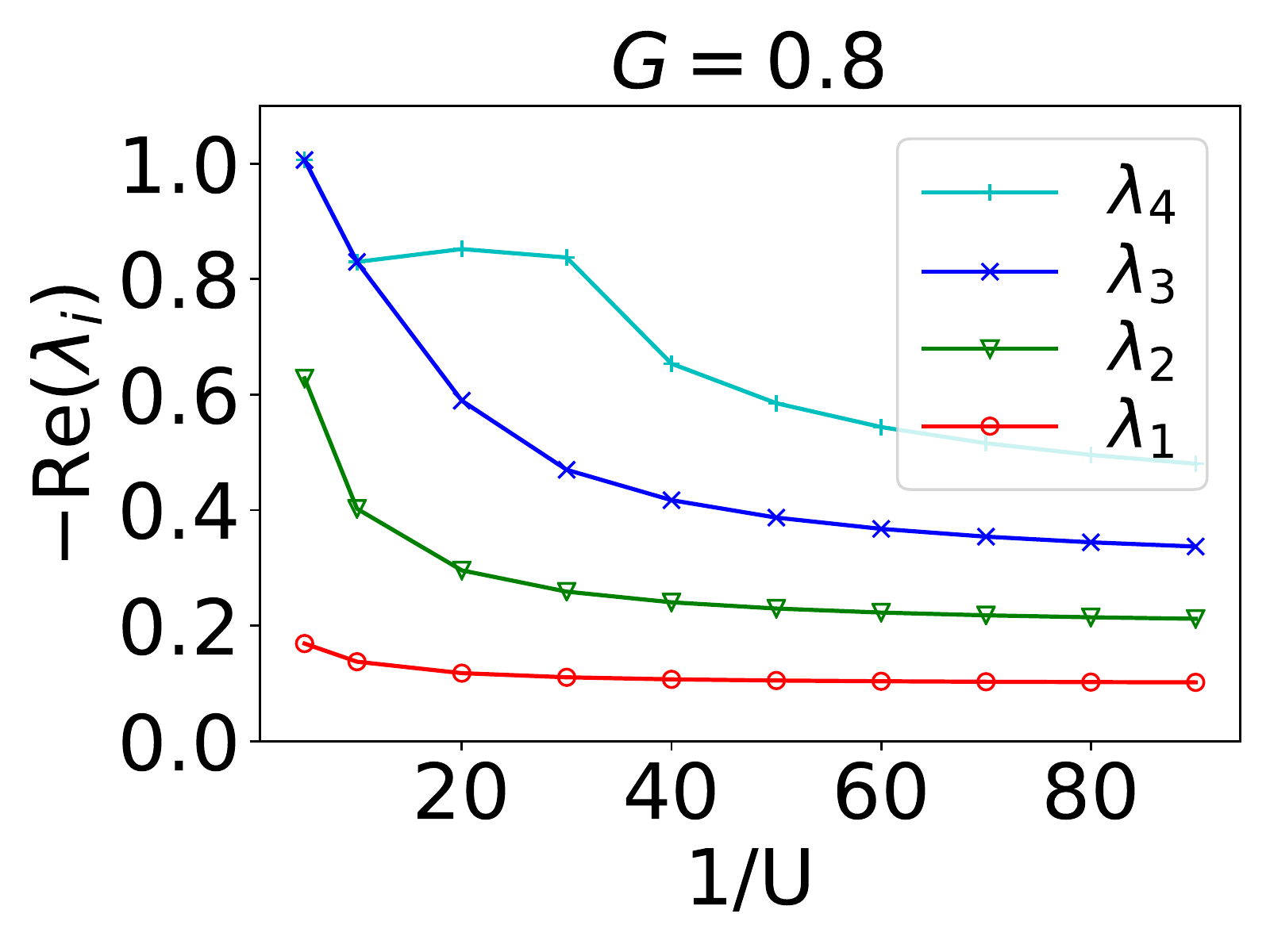}
	\includegraphics[width=0.23\textwidth]{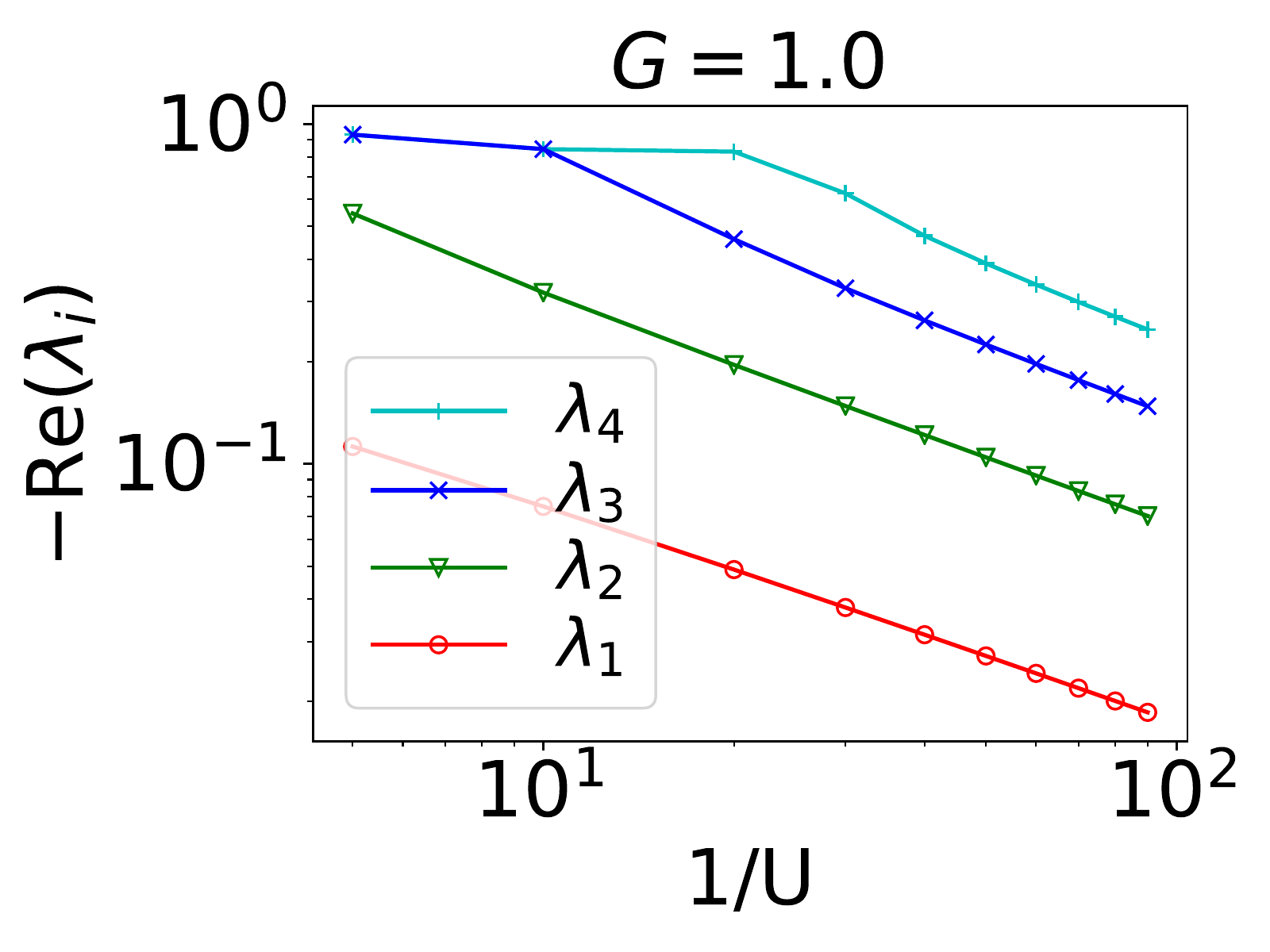}
	\put(-143,26){(a)}
	\put(-21,68){(b)}

    \includegraphics[width=0.23\textwidth]{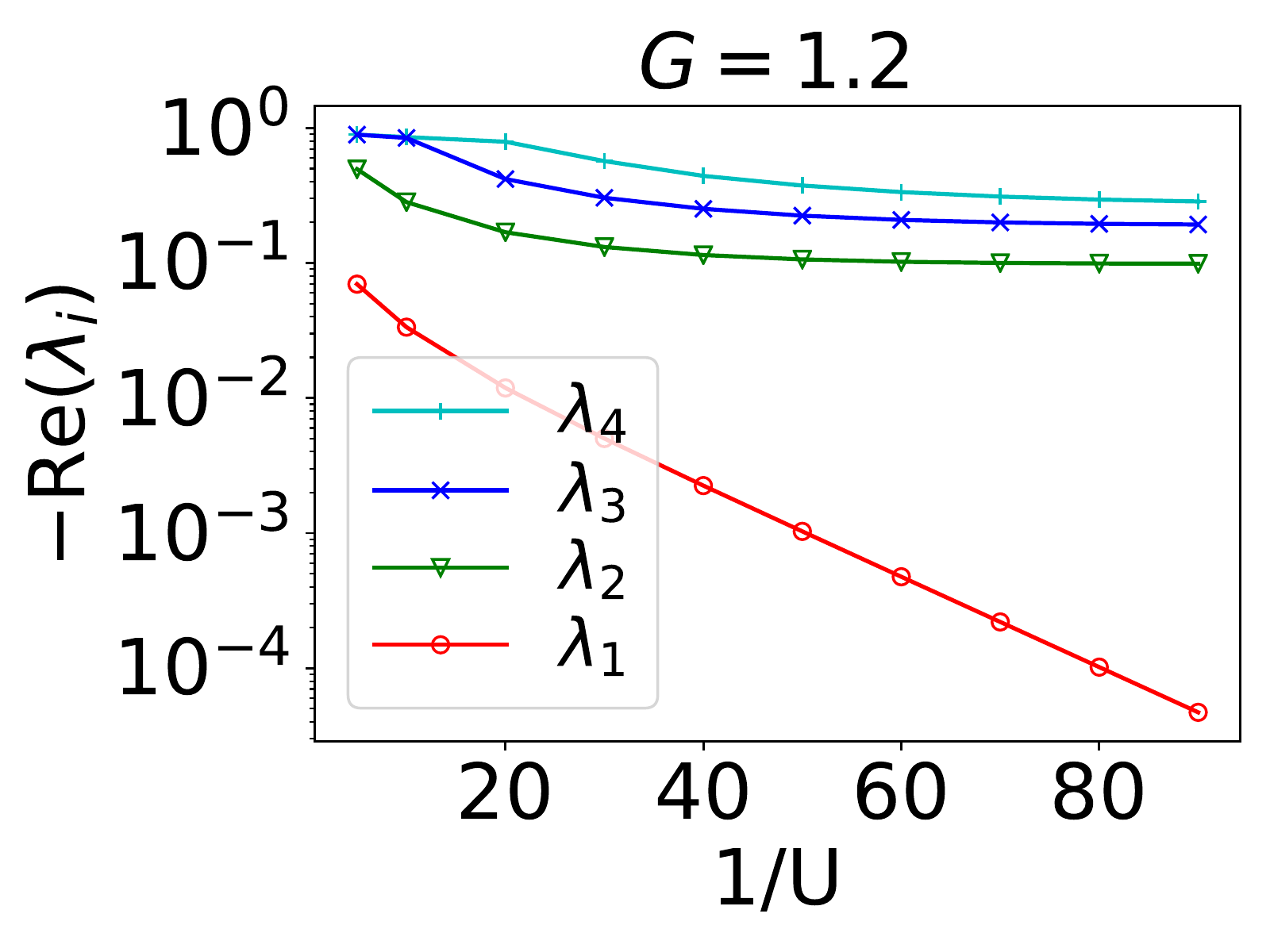}
    \includegraphics[width=0.23\textwidth]{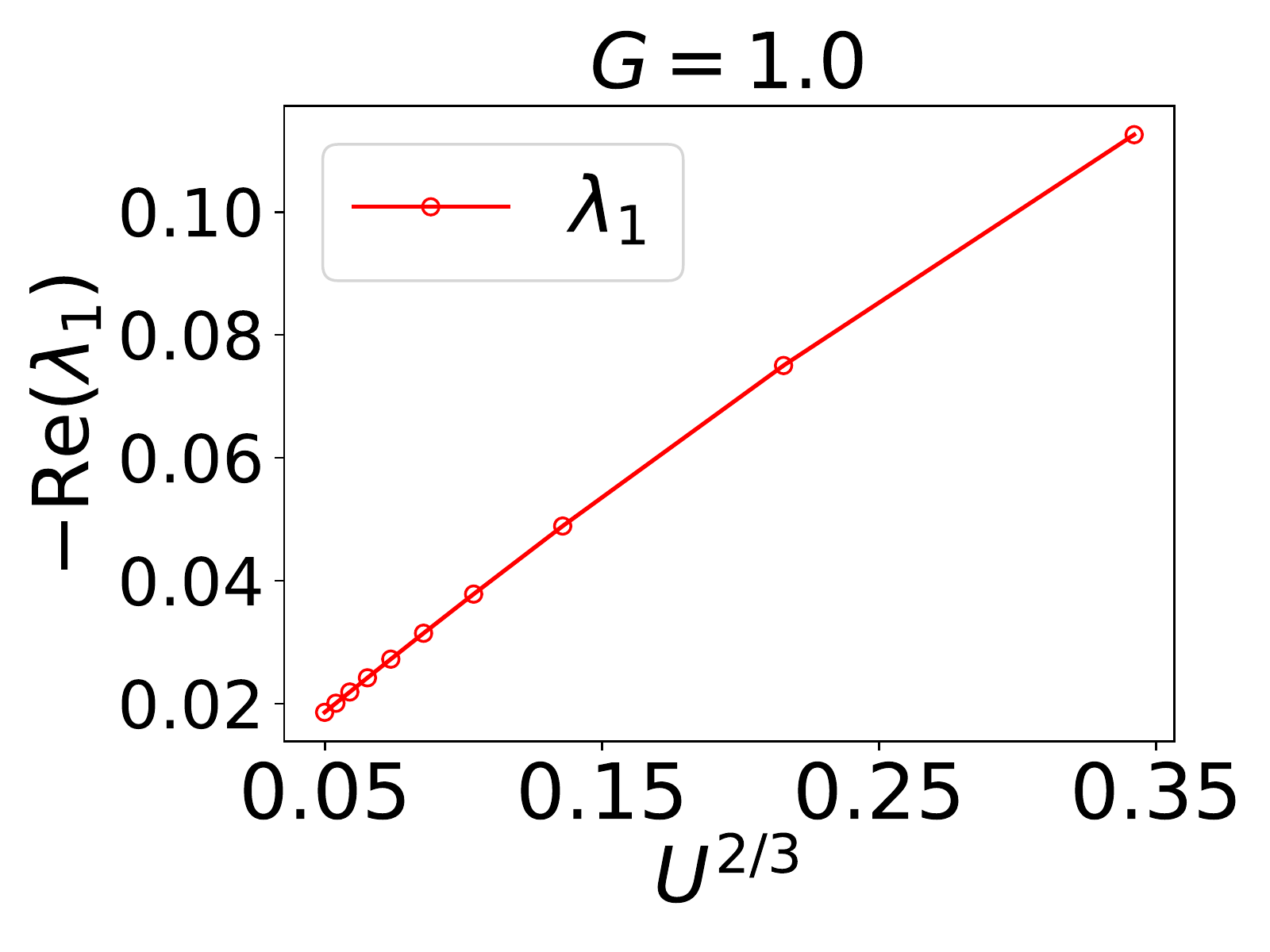}
    \put(-140,35){(c)}
	\put(-23,25){(d)}
	\caption{Finite size scaling of the real parts of highest few eigenvalues (a) below ($G=0.8$), (b) at ($G=1.0$, log-log scale), and (c) above ($G=1.2$, semi-log) the critical point. 
	The gap closes at the critical point as the decay rate of macroscopically many states approaches $0$ as a power law of $1/U$. Above the critical point, $\Re \lambda_{1}$ approaches $0$ exponentially fast, which gives a degenerate steady state manifold but the system remains gapped. (d) Shows that the gap closes as $\Delta \propto U^{2/3}$ obtained from finite size analysis.
	(Curves are ordered as in the key---the lowest curve corresponds to $\lambda_1$.)
	}
	\label{Eigenvalue}
\end{figure}

Fig.\,\ref{Eigenvalue} shows the scaling of the highest few non-zero eigenvalues as a function of $1/U$ below, at, and above the critical point.
Below the critical point, $|G|<\gamma$, where there is only one zero eigenvalue. The system is gapped.
At the critical point, the low-frequency fluctuations dominate, and the real parts of macroscopically many eigenvalues approach $0$ as a power law of $1/U$. The gap closes as $\Delta \propto U^{2/3}$, which is calculated later in section \ref{sec:QLangevin}.  Above the critical point, in addition to $\lambda_{0}=0$, $\lambda_{1}$ approaches $0$ exponentially fast, which leads to a doubly degenerate steady state manifold---i.e.\,$Z_{2}$ symmetry breaking in the thermodynamic limit. Similar to an equilibrium phase transition, the power law at the critical point signals its collective nature. Above the critical point, the exponential scaling result signals the tunneling between two metastable states.  

\begin{figure}[tb]
	\center
	\includegraphics[width=0.23\textwidth]{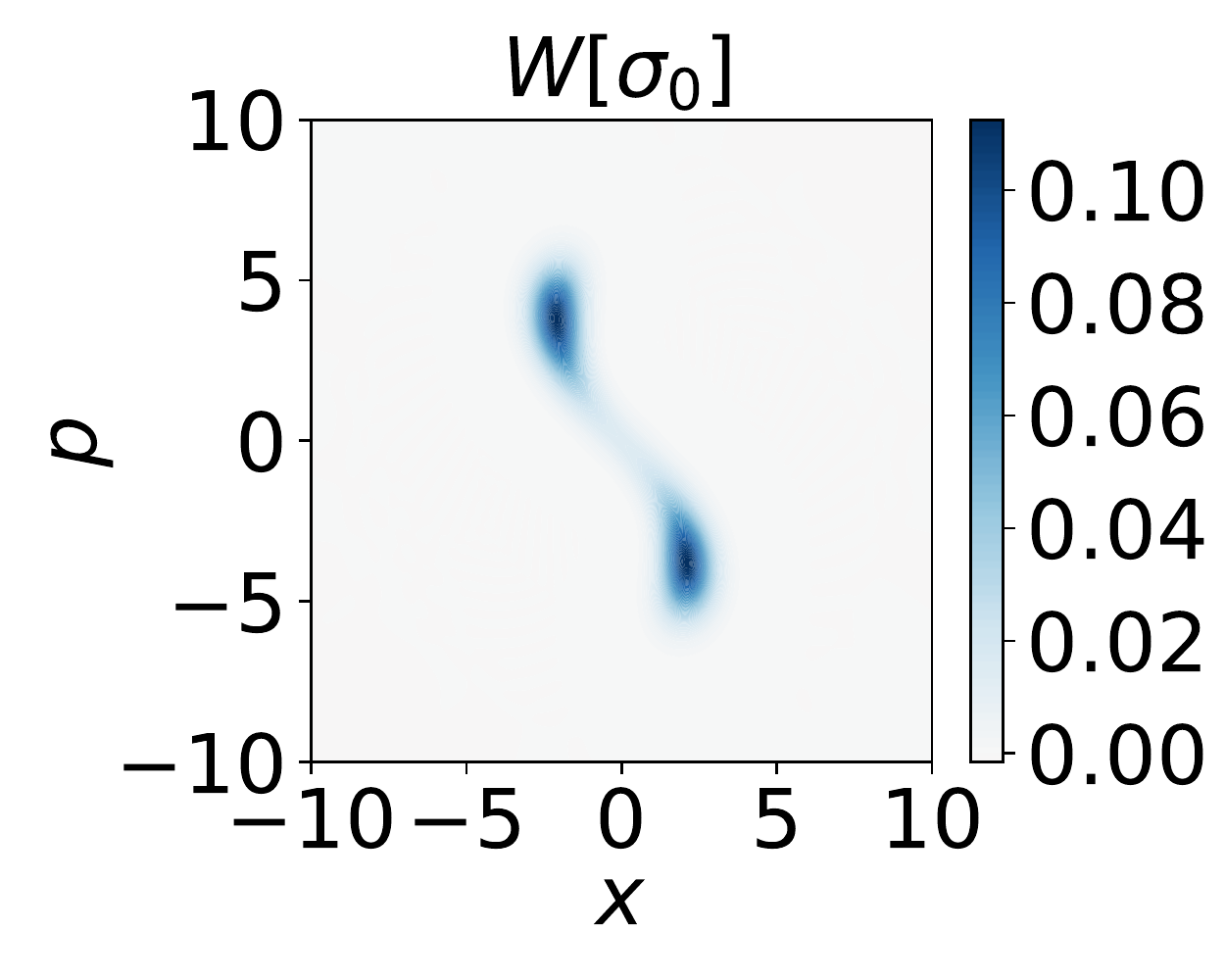}
	\includegraphics[width=0.23\textwidth]{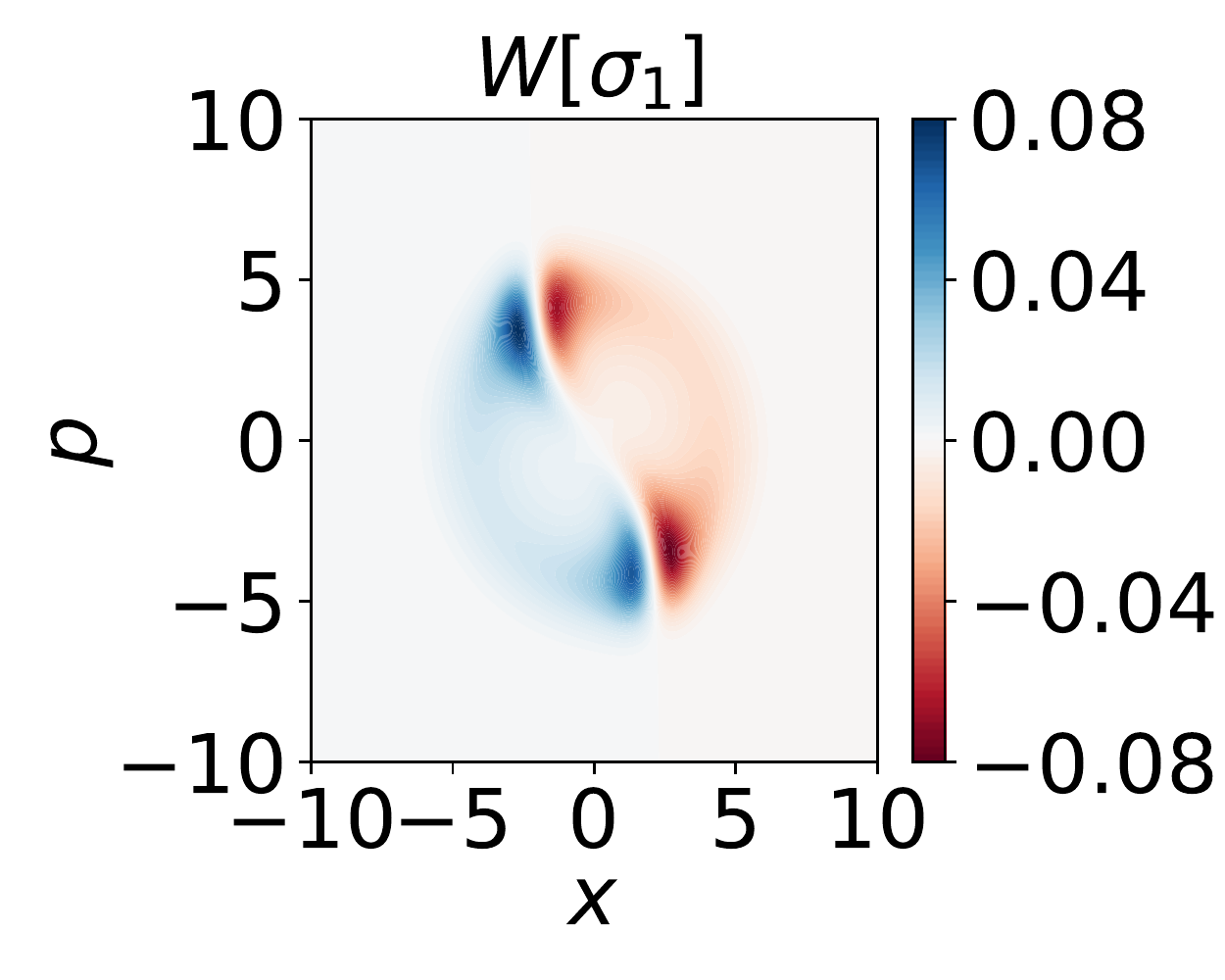}
	\put(-161,73){(a)}
    \put(-47,73){(b)}
	
	\includegraphics[width=0.23\textwidth]{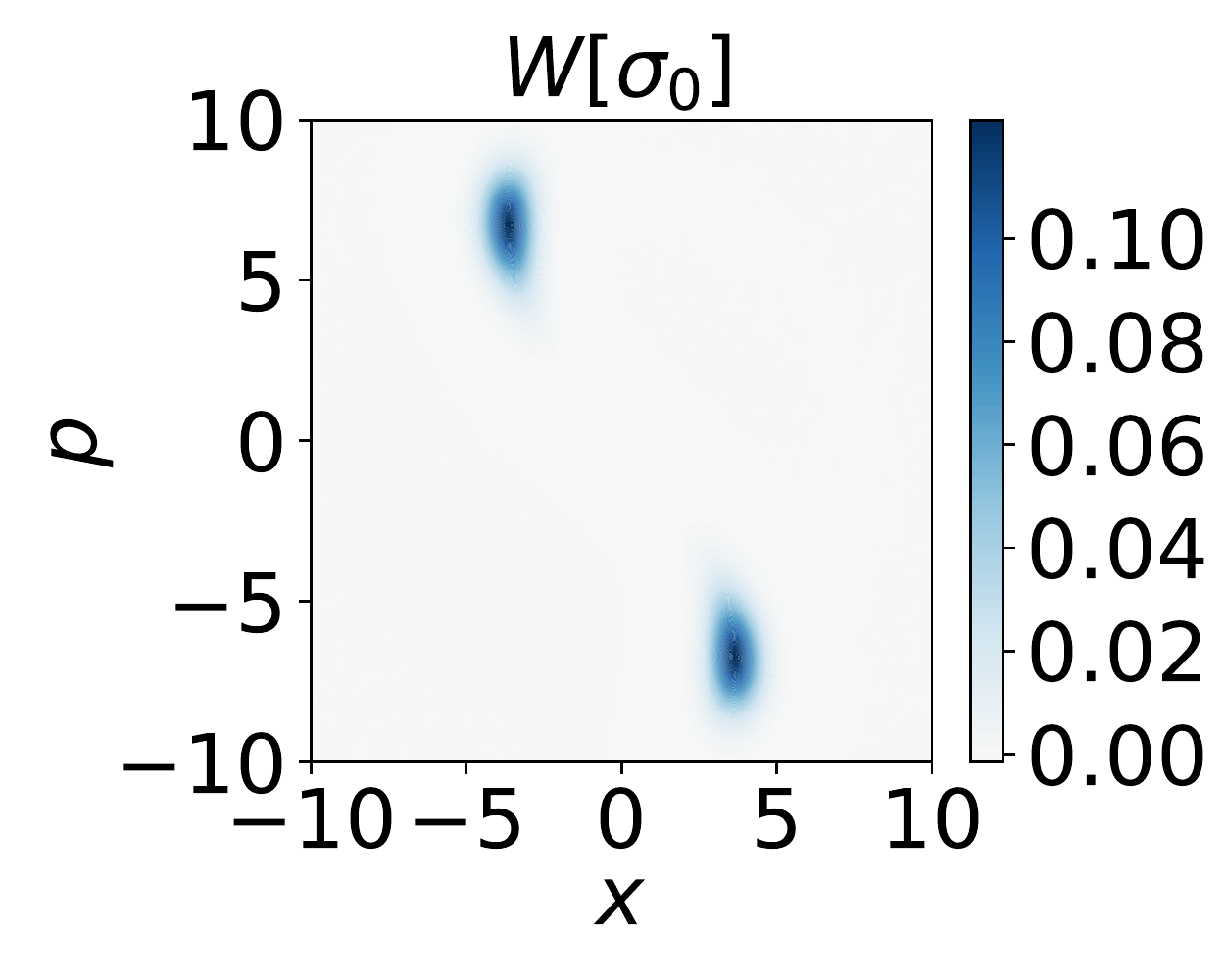}
	\includegraphics[width=0.23\textwidth]{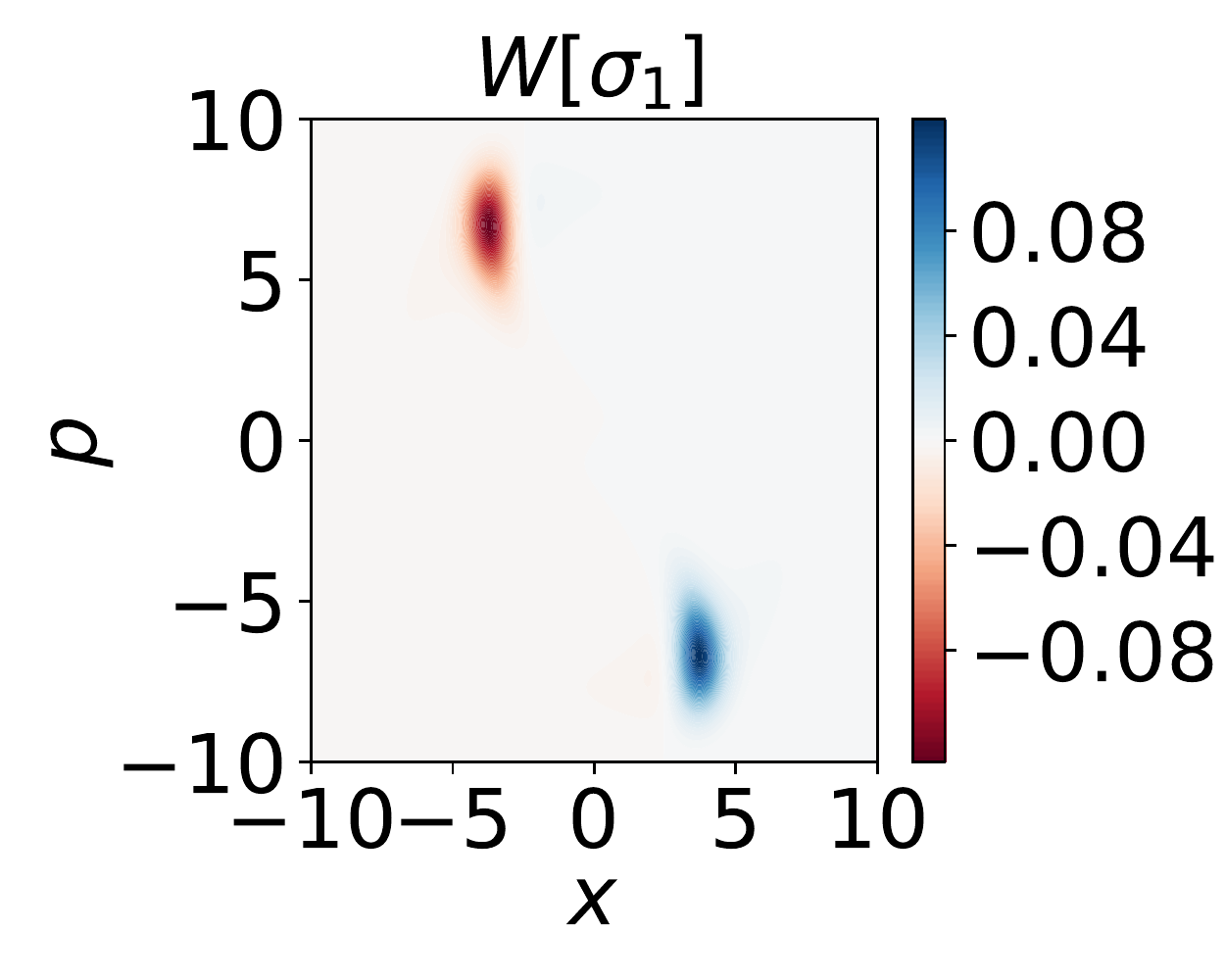}
    \put(-161,73){(c)}
    \put(-47,73){(d)}
	\caption{Wigner function of the two steady states with 0 decay rate (as $U\rightarrow0$), when driving is above critical point. $x=(a+a^{\dagger})/\sqrt{2}$ and $p=i(-a + a^{\dagger})/\sqrt{2} $. 
	Note that $W[\sigma_0]$ is symmetric about the origin while $W[\sigma_1]$ is antisymmetric. 
	Upper panels: $1/U=30$; Lower panels: $1/U=90$. (For $G=1.2$.)}
	\label{Wigner}
\end{figure}

To understand better the symmetry broken regime, we study the eigenstates $\sigma_{0}$ and $\sigma_{1}$ using their Wigner function $W[\sigma_{i=0,1}](x,p)$, with $x=(a+a^{\dagger})/\sqrt{2}$ and $p=i(-a + a^{\dagger})/\sqrt{2} $. $\sigma_{0}$ is a valid density matrix. $\sigma_{1}$ is Hermitian. However, it has only off-diagonal elements, which means it is not a valid density matrix due to lack of positivity. We normalize $\sigma_{1}$ by its trace norm such that $\Tr |\sigma_{1}|=1$. In the phase space of position and momentum, the $Z_2$ transformation amounts to $x\rightarrow -x$ and $p\rightarrow -p$. As shown in Fig.\,\ref{Wigner}, $W[\sigma_0]$ is symmetric while $W[\sigma_1]$ is antisymmetric under the $Z_2$ transformation, which means symmetry is broken in the thermodynamic limit. Both of them are squeezed due to the parametric driving. As the thermodynamic limit is reached, the mixtures $\sigma_0\pm\sigma_1$ give the two opposite amplitudes $\alpha_{s} = \pm \sqrt{n} e^{i\theta}$ respectively. 

Depending on the initial state, the steady state can break the $Z_2$ symmetry with the form $\rho = \sigma_{0} + c \sigma_1$, where $c\in \mathcal{R}$ and is constrained by the positivity of $\rho$. Notice that, unlike equilibrium cases, here there cannot be an antisymmetric steady state. This can be understood by going to the number basis. 
For an element $\ket{m}\bra{n}$ of a density matrix, the $Z_2$ transformation $\Pi = e^{i\pi a^{\dagger}a}$  yields $\Pi \ket{m}\bra{n} \Pi = (-1)^{m+n} \ket{m}\bra{n}$, which means that $\ket{m}\bra{n}$ is (anti)symmetric when $m$ and $n$ have the same (opposite) parity. Therefore, the antisymmetric states have no diagonal elements, which cannot be a valid density matrix.

\section{Keldysh Formalism}\label{sec:Keldysh}

Since ED suffers from finite size effects and MF theory completely ignores quantum fluctuations, here we use the Keldysh formalism \cite{KamenevBook2011,SiebererRPP2016} to access the thermodynamic limit while maintaining quantum fluctuations. Using the coherent state path integral of \eqref{Lindblad}, we show (see Appendix \ref{Appdix:Keldysh}) that the partition function $Z=\Tr[ \rho(t=\infty) ]$ can be written as $Z = \int \mathcal{D}[\alpha_{c}(t),\alpha_{c}^{*}(t),\alpha_{q}(t),\alpha_{q}^{*}(t)]\,e ^{i S}$ with the action 
\begin{subequations}\label{Scq}
\begin{align}
S &= \int dt \big[ (  \alpha_{c}^{*} i \frac{\partial }{\partial t} \alpha_{q} + \alpha_{q}^{*}  i \frac{\partial }{\partial t} \alpha_{c} ) - H_{\alpha}   + D_{\alpha}   \big],\\
H_{\alpha}&=\frac{U}{2}( |\alpha_{c}|^2 + |\alpha_{q}|^2 ) ( \alpha_{c}^{*}  \alpha_{q} + \text{c.c.}  ) + (\frac{G}{2} \alpha_{c}^{*}  \alpha_{q}^{*} + \text{c.c.} ), \\
D_{\alpha}&=\gamma ( i |\alpha_{q}|^2 + i \frac{1}{2} \alpha_{c}  \alpha_{q}^{*} - i \frac{1}{2} \alpha_{c}^{*}  \alpha_{q} ).
\end{align}
\end{subequations}
Here $\alpha_{c/q} =  ( \alpha_{+} \pm \alpha_{-} )/ \sqrt{2}$, where $ \alpha_{+/-}$ denotes the path in the ket/bra branch.

Then the saddle point solution $\bar{\alpha}_{\mu=c, q} (t)$ can be obtained by solving $\delta S/\delta \alpha_{\mu}=0$ and  $\delta S/\delta \alpha^{*}_{\mu}=0$, which, unsurprisingly, gives exactly the MF equation of motion \eqref{MFT} upon identifying $\bar{\alpha}_{c} = \sqrt{2} \alpha$ and $\bar{\alpha}_{q}=0$. By writing $\alpha_{\mu} = \bar{\alpha}_{\mu} + \delta \alpha_{\mu}$, the action $S$ can now be written in terms of the fluctuations $\delta \alpha_{\mu}$. The leading part, which is Gaussian, is   
\begin{equation}\label{SG}
S_{\text{G}} = \int dt \Big[ \big( \delta\alpha_{c}^{*} i \frac{\partial }{\partial t} \delta\alpha_{q} + \delta\alpha_{q}^{*}  i \frac{\partial }{\partial t} \delta\alpha_{c} \big) -  H_{\delta \alpha}^{\prime} 
+  D_{\delta \alpha} \Big],
\end{equation}
where $H_{\delta \alpha}^{\prime} \!=\! U|\bar{\alpha}_{c}|^2 ( \delta\alpha_{c}^{*}  \delta\alpha_{q} + \delta\alpha_{q}^{*} \delta\alpha_{c}  )  + (\frac{U}{2} \bar{\alpha}_{c}^2+\frac{G}{2})$ $ \delta\alpha_{c}^{*}  \delta\alpha_{q}^{*} - \text{c.c.} $ and $D_{\delta \alpha}$ is the same as $D_{\alpha}$ upon changing variables. The non-Gaussian part $S_{\text{NG}}$ contains high\-er-order fluctuations of the form $U\alpha^{*}_{c} \delta\alpha^{*}_{\mu_1} \delta\alpha_{\mu_2} \delta\alpha_{\mu_3}  $ and $U \delta\alpha^{*}_{\mu_1} \delta\alpha^{*}_{\mu_2} \delta\alpha_{\mu_3} \delta\alpha_{\mu_4}$ (see Appendix \ref{Appdix:nonGaussian}), which scale to $0$ as $\sqrt{U}$ and $U$ respectively as $U \rightarrow 0$. Therefore, in the thermodynamic limit, the non-Gaussian part is irrelevant. 

In \eqref{SG} it can be seen that an effective detuning $\omega_{d}^{\text{eff}}=- U|\bar{\alpha}_{c}|^2 = -2 \phi$ has been introduced into  $H^{\prime}_{\delta \alpha}$, which means that an effective cavity frequency $\omega_{c}^{\text{eff}} =  \omega_{c} + 2 \phi$ emerges  
\footnote{Since the term $\frac{U}{2} a^{\dagger} a^{\dagger} a a$ in the Hamiltonian can be written as $\frac{U}{2} n(n-1)$, one might expect the effective cavity frequency to be the energy cost of adding one photon $\frac{U}{2} \big((n+1)n-n(n-1)\big)=\phi$ instead of $2\phi$. This is not correct since it only considers fluctuations in photon number while in our system the fluctuating field $\alpha_{\mu}$ includes both amplitude and phase.}.  
Another collective effect is that the driving field is also shifted, becoming $G^{\text{eff}} = G + U \bar{\alpha}_{c}^2 = G + 2 U \alpha^2$. Putting in the value of $\alpha$ (for $|G|>\gamma$), we find $G^{\text{eff}} = G \big( s^2 - i s \sqrt{1- s^2} \big)$, where $s \equiv \gamma/|G| \in (0,1)$. Then the effective driving strength satisfies $|G^{\text{eff}}| = \gamma$ in the whole symmetry broken regime. 

When $|G|<\gamma$, the Kerr interaction can be ignored since $\phi= 0$, rendering the system effectively non-interacting. By solving the Heisenberg equation $\frac{d}{dt} \braket{a^{\dagger}a}=0$, we see that in the steady state
\begin{equation}\label{nonInteractingN}
\braket{a^{\dagger}a} = |G|^2/\big( 2(\gamma^2-|G|^2)\big),    
\end{equation} which is finite for $|G|<\gamma$. The same result can also be obtained using the Keldysh Green function $i g_{\text{K}}(\tau=0) =2 \braket{a^{\dagger}a} + 1 $ (see Appendix \ref{Appdix:GreenFunction}). Therefore, as $U\rightarrow 0$, the order parameter vanishes, $\phi = U \braket{a^{\dagger}a} \rightarrow 0$. Note that although $\phi=0$ below critical point, the occupation number $\braket{a^{\dagger}a}$ is not $0$. In fact, as the critical point is reached, $\braket{a^{\dagger}a}$ diverges, allowing the order parameter $\phi $ to start to be non-zero.


\section{Spectral Function}\label{sec:SpectralFunction}

Though we have shown that MF theory correctly predicts the order parameter, fluctuations actually matter and provide crucial information about the system's properties. We therefore study the spectral function
\begin{equation}
\mathcal{A}(\omega) = \frac{1}{2\pi} \big(i g_{\text{R}} (\omega) - i g_{\text{R}}^{*} (\omega) \big),
\end{equation}
where $g_{\text{R}} (\omega) \equiv -i\braket{\alpha^{*}_{q}(\omega)\alpha_{c}(\omega)} = -i\braket{\delta \alpha^{*}_{q}(\omega) \delta\alpha_{c}(\omega)} $ is the retarded Green function (see Appendix \ref{Appdix:GreenFunction}). $\mathcal{A}(\omega)$ gives an effective density of states at energy $\omega$ and thereby the probability of absorbing a weak probing signal with frequency $\omega$ [Fig.\,\ref{OrderParameter}(a)]. 

Since the quadratic fluctuations given in \eqref{SG} dominate, the spectral function can be calculated exactly (see Appendix \ref{Appdix:GreenFunction}). When $|G|<\gamma$,
\begin{equation}\label{A1}
\mathcal{A}(\omega) = \frac{1}{2\pi}  \frac{4\gamma ( 4 \omega^2 + \gamma^2 - |G|^2)}{ (-4 \omega^2 + \gamma^2 - |G|^2)^2 + 16 \gamma^2 \omega^2 },
\end{equation}
while when $|G|>\gamma$,
\begin{equation}\label{A2}
\mathcal{A}(\omega) = \frac{1}{2\pi}  \frac{\gamma }{ (\omega - 2\phi)^2 + \dfrac{\gamma^2 \omega^2}{(\omega + 2\phi)^2} }. 
\end{equation}

\begin{figure}[tb]
	\center
	\includegraphics[width=0.23\textwidth]{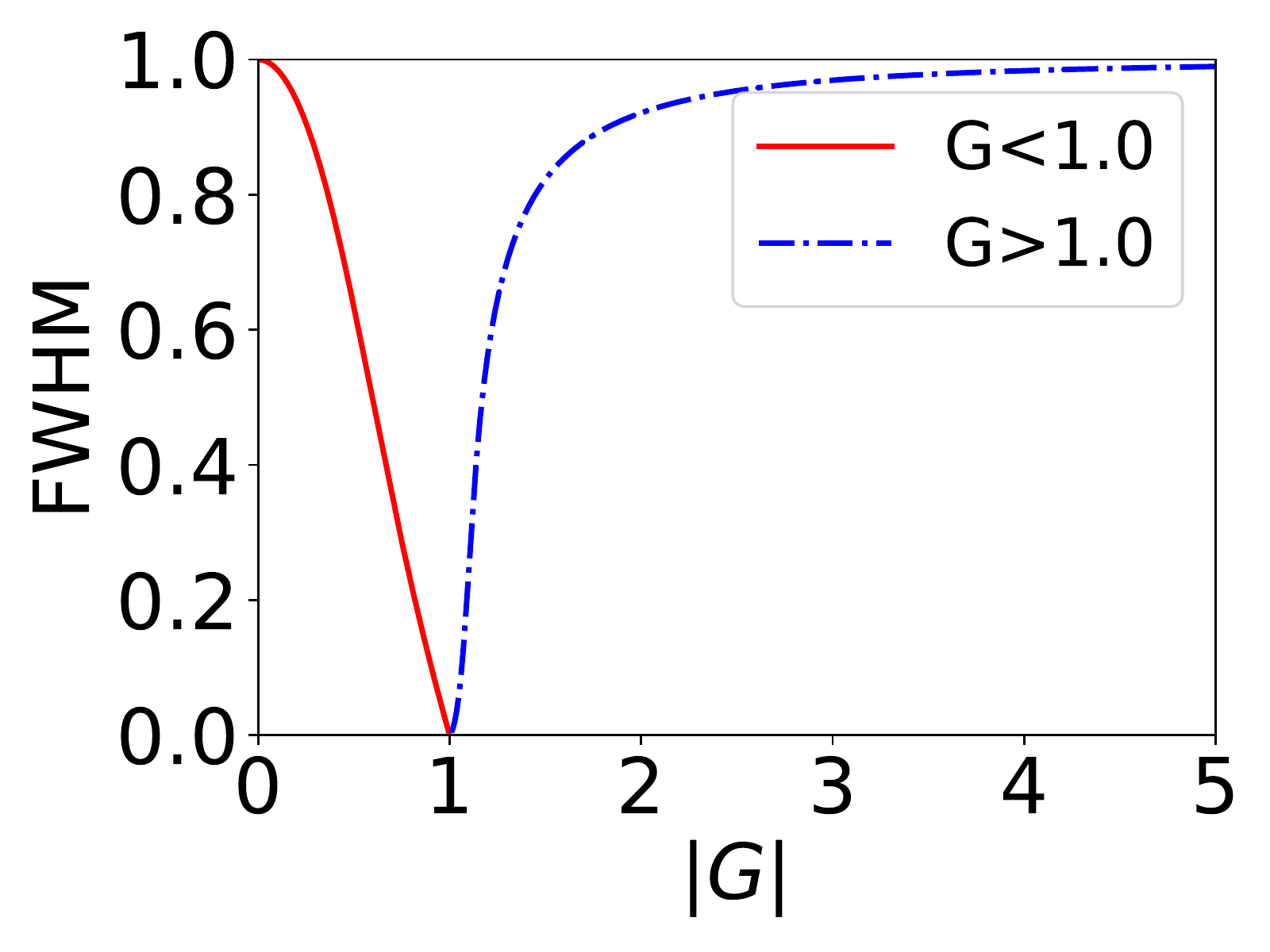}
	\includegraphics[width=0.23\textwidth]{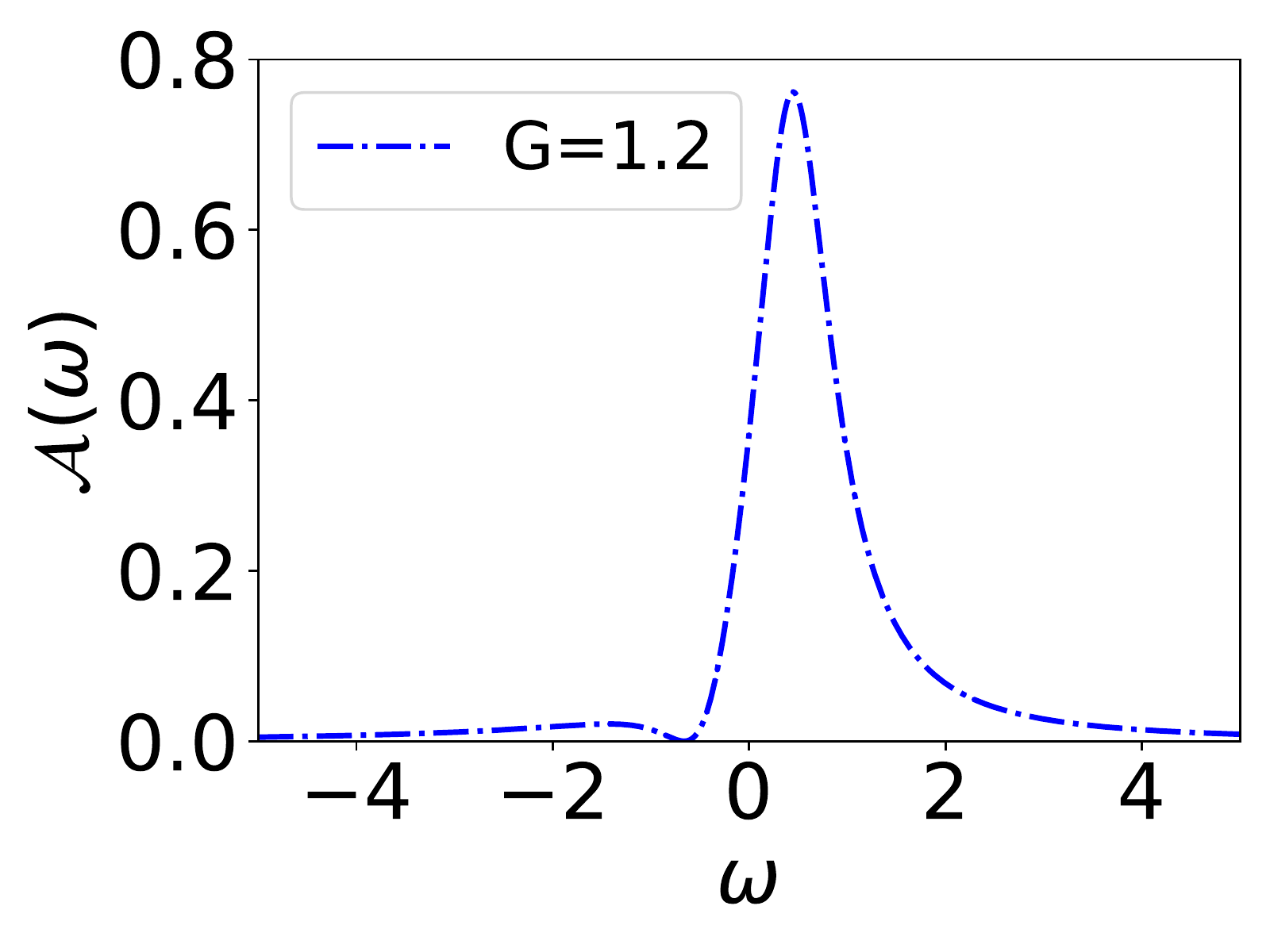}
\put(-140,25){(a)}
\put(-90,60){(b)}

	\includegraphics[width=0.23\textwidth]{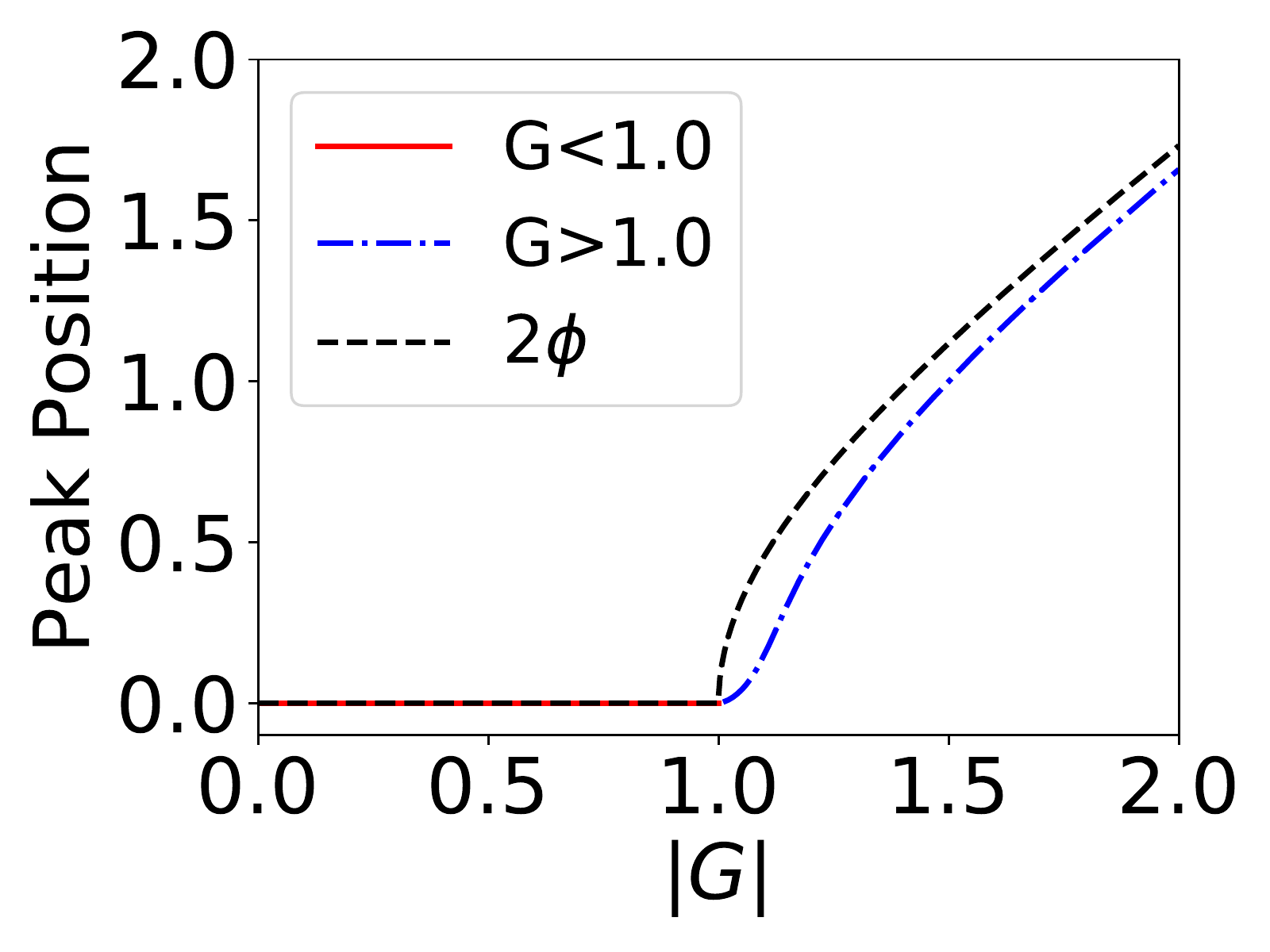}
	\includegraphics[width=0.23\textwidth]{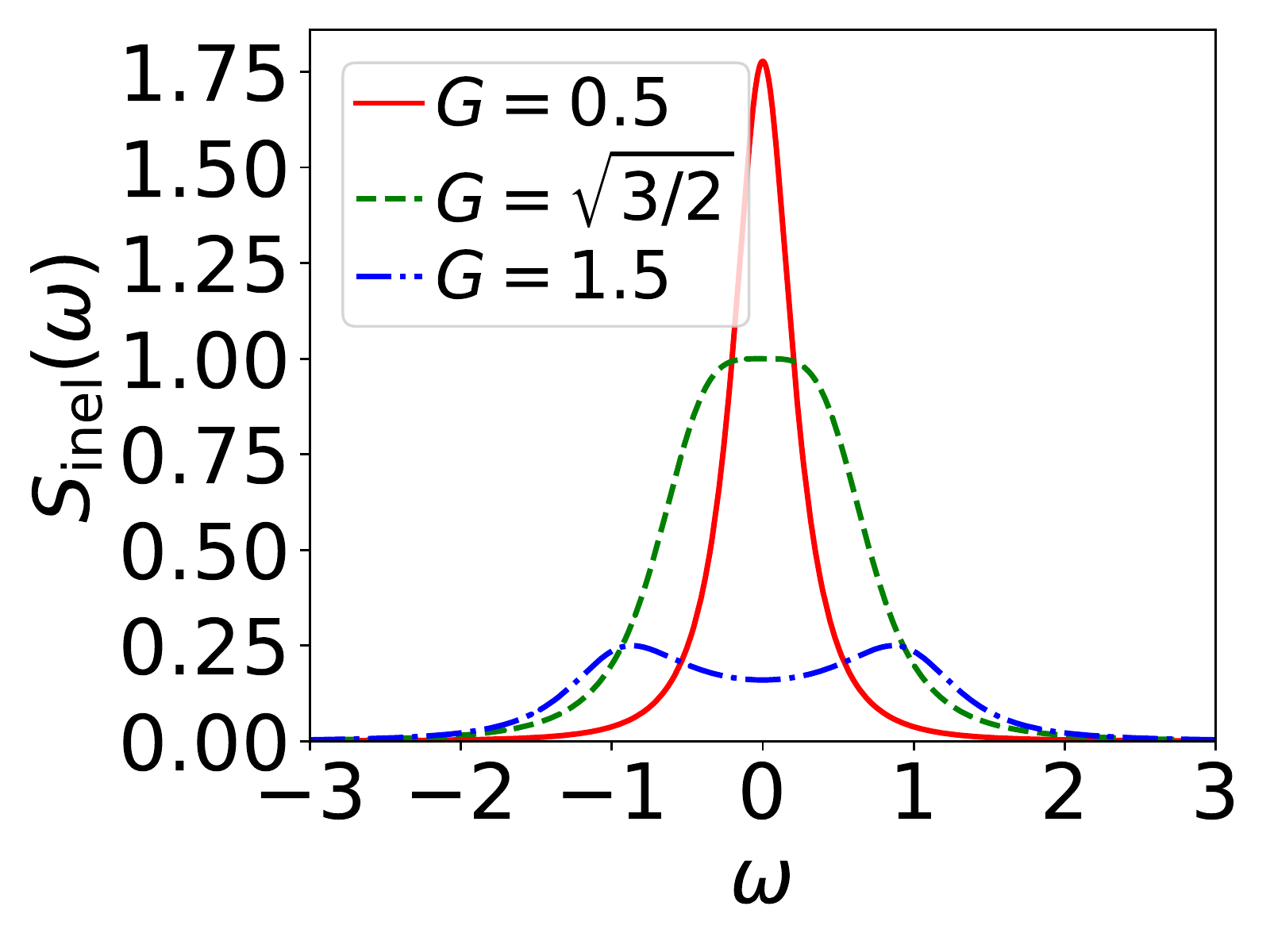}
\put(-145,25){(c)}
\put(-20,75){(d)}
	\caption{(a) FWHM of the absorption spectrum $\mathcal{A}(\omega)$ as a function of the driving strength $|G|$. (b) $\mathcal{A}(\omega)$ at $|G|=1.2$, which has an asymmetric line profile. (c) Position of the peak of $\mathcal{A}(\omega)$ as a function of $|G|$. Dashed line is $2\phi$, which is also the position of $\mathcal{A}(\omega)=0$ in the symmetry broken regime. (d) Inelastic power spectrum of emitted photons $S_{\text{inel}}(\omega)$.   }
	\label{Aw}
\end{figure}

Below the critical point, the absorption spectrum, given by \eqref{A1}, is an even function centered at $0$. When $|G|=0$, $\mathcal{A}(\omega)$ is a Lorentzian, as expected, with full width at half maximum (FWHM) equal to $\gamma$. As the critical point is reached,  the FWHM  decreases monotonically to $0$ [Fig.\,\ref{Aw}(a)], which means the low-energy fluctuations dominate as expected for quantum criticality. The zero-width peak corresponds to fluctuations with infinitely long lifetime, leading to long-range time correlations. This criticality is consistent with the spectrum of the Lindbladian [Fig.\,\ref{Eigenvalue}(a)]: a macroscopically large number of eigenstates with zero decay rate in the thermodynamic limit.  

\begin{figure}[tb]
	\center
	\includegraphics[width=0.23\textwidth]{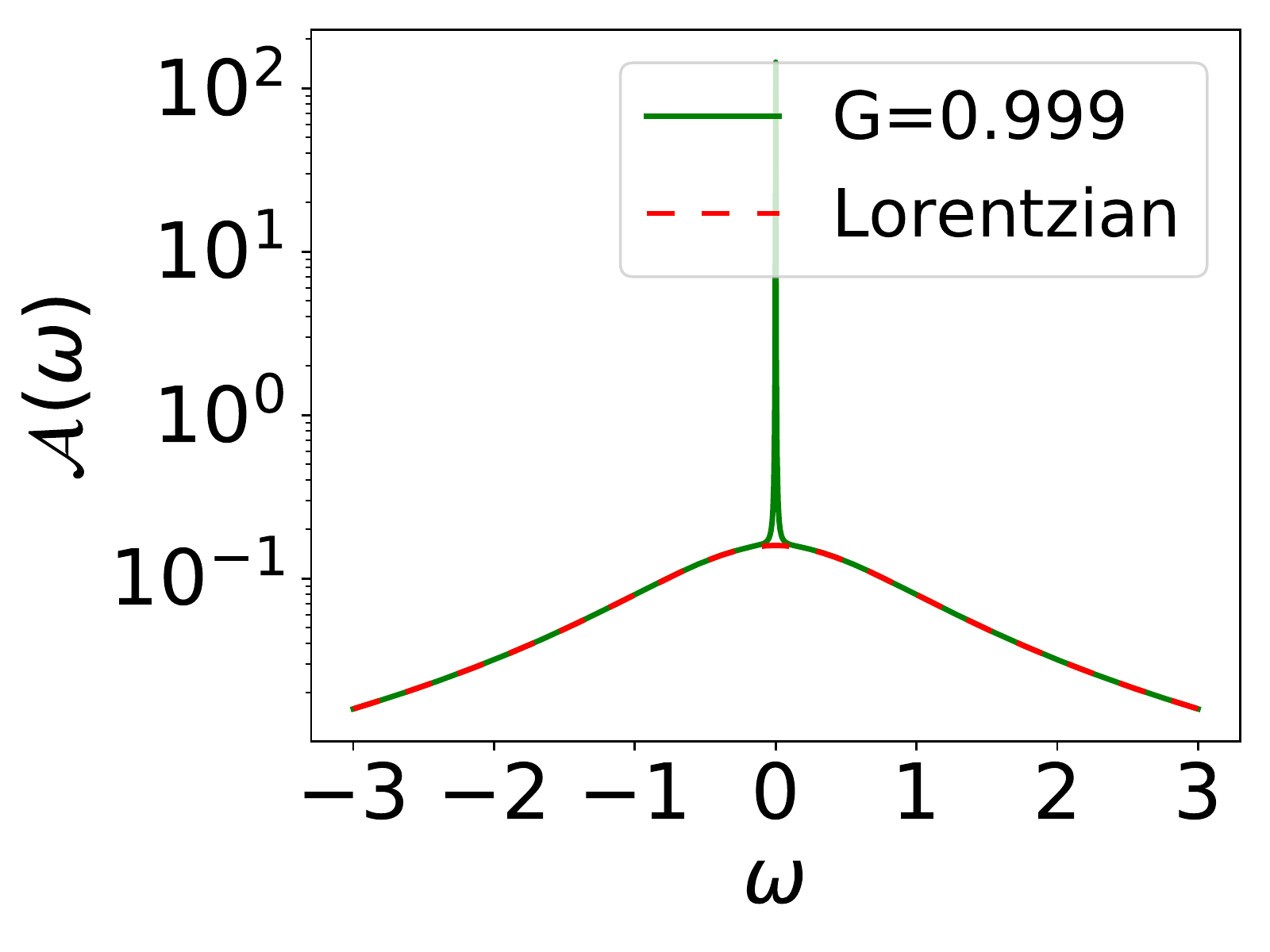}
	\includegraphics[width=0.23\textwidth]{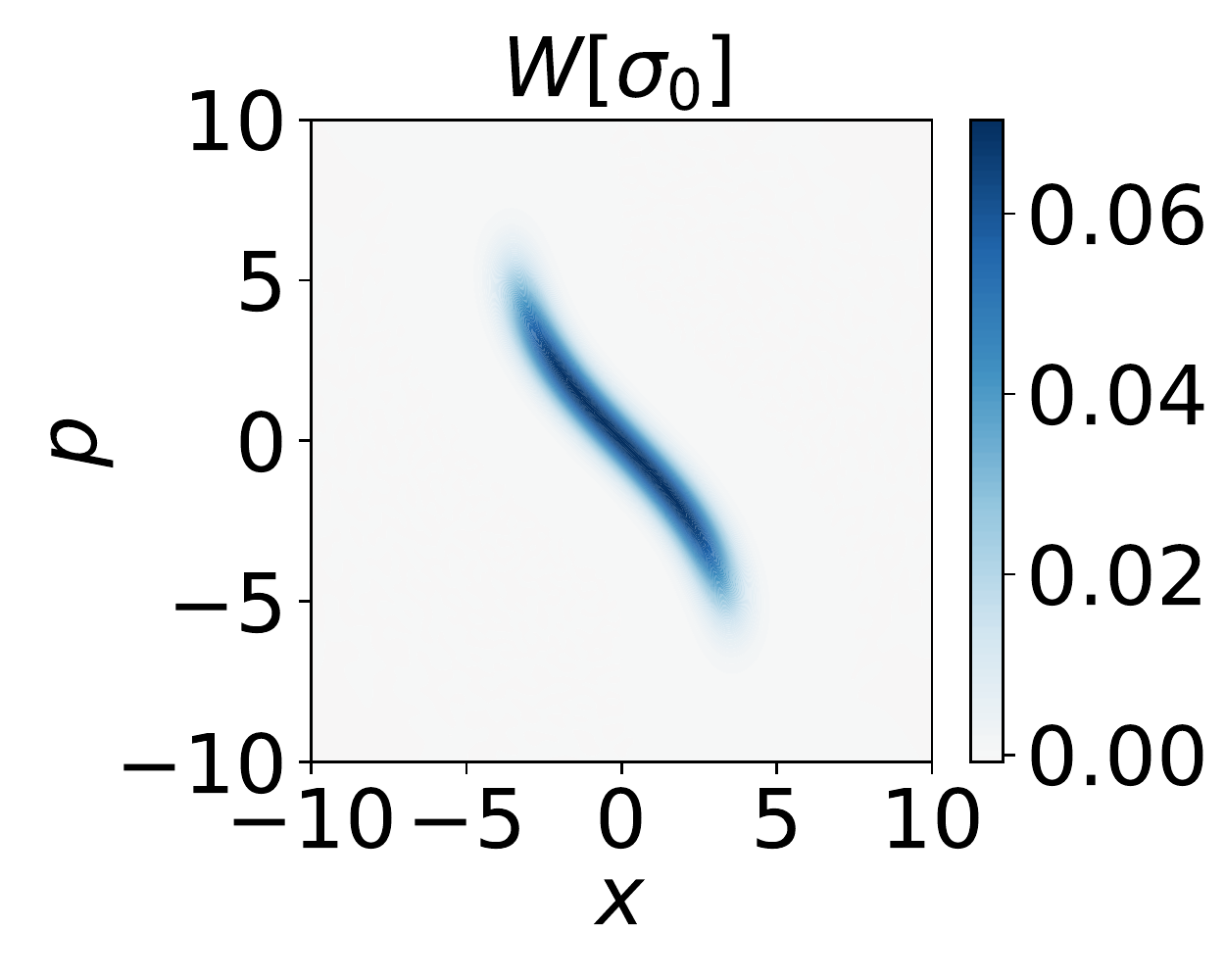}
\put(-203,73){(a)}
\put(-83,73){(b)}
	\caption{(a) Spectral function as the critical point is reached. For $G=0.999$, Eq.\,\eqref{A1} is used while the Lorentzian base is obtained from Eq.\,\eqref{Ac}. (b) Wigner function of the steady state when $G=1$ and $1/U=90$. The quadrature  $X_{2} = (a e^{-i\pi/4}+ a^{\dagger} e^{i\pi/4})$ becomes perfectly squeezed as $U\rightarrow0$. }
	\label{CriticalA}
\end{figure}

To get the spectrum at the critical point, we can see that, for $|G|<\gamma$,
\begin{equation}\label{gRtau}
i g_{\text{R}} (\tau) - i g_{\text{A}} (\tau) = \frac{1}{2}  ( e^{-\frac{\gamma-|G|}{2}|\tau|} + e^{-\frac{\gamma+|G|}{2}|\tau|}),
\end{equation}
whose Fourier transform gives \eqref{A1}, which is a sum of two Lorentzian functions.
As the critical point is reached,
\begin{equation}
i g_{\text{R}} (\tau) - i g_{\text{A}} (\tau) = \frac{1}{2}  ( 1 + e^{-\gamma|\tau|}),
\end{equation}
which gives the spectral function at critical point
\begin{equation} \label{Ac}
\mathcal{A}(\omega) = \frac{1}{2} \left( \delta (\omega) + \frac{1}{\pi}  \frac{\gamma }{ \omega^2 + \gamma^2  } \right),  
\end{equation}
which is an equal weight mixture of a Dirac delta function and a Lorentzian with width $=2\gamma$. Indeed, the states with 0 decay rate give a delta function while those with non-zero decay rate give a Lorentzian.  Fig.\,\ref{CriticalA}(a) shows the spectral function obtained with Eq.\,\eqref{A1} for a case very close to the critical point, and shows furthermore that it agrees with Eq.\,\eqref{Ac}.  

Above the critical point, \eqref{A2} shows that the order parameter $\phi$ and Kerr interaction come into play. $G$ does not appear explicitly because $|G^{\text{eff}}|=\gamma$. $\mathcal{A}(\omega)$ is not symmetric anymore, and its peak shifts away from zero. As an example, $\mathcal{A}(\omega)$ at $|G|=1.2$ is shown in Fig.\,\ref{Aw}(b). $\mathcal{A}(\omega)$ reaches its maximum at a positive frequency, which is $\approx 2 \phi$ for large $G$ as shown in Fig.\,\ref{Aw}(c). 

When $|G| \gg \gamma$, $\mathcal{A}(\omega)$ becomes a Lorentzian again. To see this, we expand $\mathcal{A}(\omega)$ around $2\phi$ and set $\omega = 2 \phi + \delta \omega$:
\begin{equation}
\begin{split}\label{AwLargeG}
\mathcal{A}(\omega)  \approx \frac{1}{\pi}  \frac{\gamma/2 }{ \delta\omega^2 + \big(\gamma(1+\delta\omega/4\phi + \dots )/2\big)^2  }.
\end{split}
\end{equation}
Thus, when $|G|\gg \gamma$, $\delta\omega \sim \gamma \ll \phi$, which makes \eqref{AwLargeG} a Lorentzian with FWHM $\rightarrow\gamma$, as for $G=0$. This can be understood by analyzing the action $\eqref{SG}$ using $H_{\delta \alpha}^{\prime} = -\omega_{d}^{\text{eff}} ( \delta\alpha_{c}^{*}  \delta\alpha_{q} + \text{c.c.}  )  + \frac{G^{\text{eff}}}{2} \delta\alpha_{c}^{*}  \delta\alpha_{q}^{*} + \text{c.c.} $. When $|G|\gg \gamma$,  $|G^{\text{eff}}|=\gamma$, which is negligible compared to the large detuning $\omega_{d}^{\text{eff}} = -2\phi \approx -|G|$. Therefore, for strong driving, the theory looks like that of a cavity with frequency $2\phi$ coupled to an external environment without any driving---the same as for $G=0$ (with the cavity frequency shifted).

Notice the striking feature $\mathcal{A}(\omega=-2 \phi)=0$ exactly, which originates from the squeezing of the Lorentzian peak by the parametric driving. Therefore for weak probing light of frequency $\omega_{c}-2\phi$, there is no absorption. 
Measurement of the frequency with zero absorption thus provides a direct measure of the order parameter $\phi$, complementary to measuring the occupation number.

\section{Power spectrum of the output field}\label{sec:PowerSpectrum}

Fluctuations are crucial for other 
physically important quantities, such as the power spectrum of the output field $S(\omega)=$ 
$\int d\omega e^{i\omega \tau}\braket{a^{\dagger}_{\text{out}}(t)a_{\text{out}}(t+\tau)}$, 
which gives the energy dis\-tribution of photons emitted through the dissipation channel [Fig.\,\ref{OrderParameter}(a)]. From input-output theory \cite{WallsMilburnBook,CarmichaelBook2008}, we know that 
$a_{\text{out}} (t) = \sqrt{\gamma} a (t)$. In the Keldysh formalism, using the operator expressions of Green functions, we find
\begin{equation}
\begin{split}
S(\omega)=i \frac{\gamma}{2} \Big( g_{K}(\omega) + g_{A}(\omega)  - g_{R}(\omega) \Big)
\end{split}
\end{equation}
 where $g_{\text{K}}(\omega)=-i\braket{\alpha^{*}_{c}(\omega)\alpha_{c}(\omega)}$ and $g_{\text{A}}(\omega) = g_{\text{R}}^{*}(\omega)$ $= -i\braket{\alpha^{*}_{c}(\omega)\alpha_{q}(\omega)}$ are the Keldysh and advanced Green function, respectively. When written in terms of fluctuations, we see that $i g_{\text{K}}(\omega)= |\bar{\alpha}_{c}|^2 \delta(\omega) + \braket{\delta\alpha_{c}^{*}(\omega)\delta\alpha_{c}(\omega)}$, where the first term with $|\bar{\alpha}_{c}|^2 =2 \phi/U$ is due to elastic scattering. Then the inelastic scattering power spectrum is $S_{\text{inel}} (\omega) = \gamma \braket{\delta\alpha_{c}^{*}(\omega)\delta\alpha_{c}(\omega)}/2 - \gamma \pi \mathcal{A}(\omega) $.  
 
Using the expressions for the Green functions in Appendix \ref{Appdix:GreenFunction}, we find 
\begin{equation} \label{Sw}
S_{\text{inel}} (\omega) = \begin{cases} 
\dfrac{4\gamma^2 |G|^2}{ (-4 \omega^2 + \gamma^2 - |G|^2)^2 + 16 \gamma^2 \omega^2 } & |G|<\gamma, \\[0.3cm]
 \dfrac{\gamma^4}{4} \dfrac{1}{(\omega^2-4\phi^2)^2+\gamma^2\omega^2} & |G|>\gamma.
\end{cases}
\end{equation}
Unlike the absorption spectrum given by the spectral function, the power spectrum is always symmetric because processes with emitted frequencies $\omega_{c}+\omega$ and $\omega_{c}-\omega$ have the same amplitude by energy conservation \cite{DrummondJPA1980}. The linewidth goes to $0$ at the critical point, like the spectral function. Using the same approach as in obtaining Eq.\,\eqref{Ac}, we see that for $|G| \lesssim \gamma $,
\begin{equation}
S_{\text{inel}} (\omega) = \frac{\gamma^2}{4\omega^2+\Delta_{G}^2},
\end{equation}
where $\Delta_{G}=\gamma - |G|$. The power spectrum has a $1/\omega^2$ divergence as the critical point is reached. 
Since the Kerr interaction can be ignored below the critical point, the system acts like 
a degenerate parametric oscillator \cite{WallsMilburnBook,CarmichaelBook2008} in that regime. 
From the known squeezing spectra for the latter, we conclude that the suppression of fluctuations in our system results from the squeezing of the quadrature 
$X_{2} \equiv (- i a e^{i\theta}+i a^{\dagger} e^{-i\theta})$ by the parametric driving, where $\theta = \theta_{G}/2 $ for $G= i |G| e^{-i \theta_{G}}$ (see Eq.\,\eqref{phasetheta}). This can also be seen from the MF solution of the steady state amplitude $\alpha_{s}$ at the critical point. At the critical point, the quadrature $X_2$ becomes perfectly squeezed as seen in Fig.\,\ref{CriticalA}(b). Fluctuations along its orthogonal direction $X_1$ therefore produce a zero-width peak in both the spectral function and power spectrum. This can also be seen later in Eq.\,\eqref{QLangevinEq}. In the symmetry broken regime, as $|G|$ increases two peaks 
arise from the effective detuning
when $|G|>\sqrt{3/2}\gamma$ such that the detuning is larger than the peak width [Fig.\,\ref{Aw}(d)].

\section{Quantum Langevin Equation and Critical Scaling} \label{sec:QLangevin}

The quantum Langevin equation is a noisy equation of motion for operators which, in contrast with the MF theory, gives the exact dynamics of operators in the sense that all orders of correlation can be calculated exactly. It can be derived using the Heisenberg equation of motion and a noise operator \cite{GardinerZollerBook}. Here, in order to show its connection with Keldysh formalism, we derive it using a Hubbard-Stratonovich transformation \cite{SiebererRPP2016}. Following the analysis in \cite{TorrePRA2013}, we can then compute the scaling exponents based on the low-frequency expansion of the quantum Langevin equation at the critical point. Although scaling upon approaching the critical point can also be obtained with the Keldysh formalism, the quantum Langevin equation is particularly useful for finite-size analysis at the critical point. 

\subsection{Quantum Langevin equation from Keldysh action}

As we know from the MF theory, for $|G|\lesssim \gamma$, the steady state amplitude $\alpha_{s}$ is squeezed in the direction $(- i  \alpha_{s} e^{i\theta_{G}/2}+i \alpha_{s}^{*} e^{-i\theta_{G}/2})$ while amplified in the orthogonal direction. Without of loss of generality, we set $G=ig$ with $g$ being real i.e.\ $\theta_{G}=0$. We can define a set of real variables $x_{c/q} = (\alpha_{c/q}+\alpha_{c/q}^{*})/\sqrt{2}$ and $p_{c/q} = (-i\alpha_{c/q}+i\alpha_{c/q}^{*})/\sqrt{2}$. Then $x_{c}$ is amplified while $p_{c}$ is squeezed. Using this coordinate transformation in \eqref{SGmatrix}, we get the Keldysh action in the $xp$ basis:
\begin{equation}
\begin{split}
S_{xp} &= \int d\omega\, \big[\, i\frac{\gamma}{2} x_{q}^{*}(\omega) x_{q}(\omega) + i\frac{\gamma}{2} p_{q}^{*}(\omega) p_{q}(\omega) \\ 
&+ \frac{1}{2} (\gamma-g - 2i\omega) x_{c}(\omega) p_{q}^{*}(\omega)  \\ 
&- \frac{1}{2} (\gamma+g - 2i\omega) p_{c}(\omega) x_{q}^{*}(\omega)  \,\big],
\end{split}
\end{equation}
where $x_{c/q}(\omega)^{*} = x_{c/q}(-\omega)$ and $p_{c/q}(\omega)^{*} = p_{c/q}(-\omega)$ as they are real variables in real time.

Introducing a Hubbard-Stratonovich transformation \cite{SiebererRPP2016}, we know
\begin{equation}
\begin{split}
e^{-\frac{\gamma}{2} \int d\omega p_{q}^{*}(\omega) p_{q}(\omega)} \propto \int \mathcal{D}[f_{x}]  e^{-\frac{2}{\gamma} \int d\omega f_x(\omega)^{*}f_x(\omega)} \\ \times e^{i \int d\omega 2  f_x(\omega)p_{q}^{*}(\omega)},
\end{split}
\end{equation}
where $f_x(t)$ can be interpreted as real Gaussian white noise with $\braket{f_x^{*}(\omega) f_{x}(\omega^{\prime})} =  \delta(\omega-\omega^{\prime}) \frac{\gamma}{4}$ and $f_x^{*}(\omega) = f_x(-\omega)$. Thus an equivalent expression for the partition function $Z$ is obtained  in which the Keldysh action $S$ depends on $p_{q}(\omega)$ only linearly. Similarly, we can introduce another Gaussian noise $f_p(\omega)$ with the same properties to eliminate the quadratic term $\frac{\gamma}{2}x_{q}^{*}(\omega) x_{q}(\omega)$. Since now the action depends linearly on both $x_{q}(\omega)$ and $p_{q}(\omega)$, they can be directly integrated out. Then the partition function becomes 
\begin{equation}\label{QLangevinZ}
\begin{split}
Z =& \int \mathcal{D}[x_{c},p_{c},f_{x},f_{p}]\, e^{-\frac{2}{\gamma} \int d\omega \left(f_x(\omega)^{*}f_x(\omega) + f_p(\omega)^{*}f_p (\omega)\right)} \\
\times& \delta\left( \frac{1}{2} (\gamma-g - 2i\omega) x_{c}(\omega) + 2f_{x}(\omega)  \right) \\
\times& \delta\left( \frac{1}{2} (\gamma+g - 2i\omega) p_{c}(\omega) + 2f_{p}(\omega)  \right),
\end{split}
\end{equation}
which shows that $x_{c}$ and $p_{c}$ satisfy exactly the quantum Langevin equation 
\begin{equation}\label{QLangevinEq}
\begin{cases} \frac{d}{dt} x_{c}(t) + \frac{1}{2} (\gamma-g) x_{c}(t) + 2 f_{x}(t) = 0,  \\[0.1cm]
\frac{d}{dt} p_{c}(t) + \frac{1}{2} (\gamma+g) p_{c}(t) + 2 f_{p}(t) = 0.
\end{cases}
\end{equation}
Using it, all orders of correlations of $x_{c}(t)$ and $p_{c}(t)$ can be calculated exactly, since the partition function \eqref{QLangevinZ} is equivalent to the original one. They are in the same form as the Langevin equations that describe a massless particle moving in a harmonic potential, subject to friction and thermal noise $2f$ \cite{TorrePRA2013,TongKineticTheoryNotes}. We can define an effective temperature $T_{\text{eff}} =  \gamma/2$, and thus find the steady state distribution of $x_{c}$  
\begin{equation}\label{thermalPx}
P(x_{c}) = \frac{1}{\mathcal{N}} e^{- F(x_{c})/T_{\text{eff}}},    
\end{equation}
where $\mathcal{N}$ is a normalization factor and $F(x_{c}) =  \frac{1}{4} (\gamma-g) x_{c}^2$ is the effective free energy. Similarly, $P(p_{c})$ can be obtained from $\tilde{F}(p_{c}) =  \frac{1}{4} (\gamma+g) p_{c}^2$. Notice that Eq.\,\eqref{thermalPx} gives an interesting picture of the criticality: as the critical point is reached [i.e.\ $(\gamma-g)\rightarrow0$], $F(x_{c})$ becomes a flat potential and so the finite-temperature occupation $\braket{x_{c}^2}$ diverges. 

\subsection{Critical exponent of occupation number }\label{ScalingN}

As we saw above, the occupation number $n$ diverges as the critical point is reached. Let the scaling be $n \sim (\gamma-g)^{-\nu_{n}}$, where $\nu_{n}$ is the critical exponent of the occupation number. Here we only consider critical exponents from below. Exponents from above are generally the same except in rare cases \cite{LeonardPRL2015}. As shown in Appendix \ref{Appdix:ExponentAbove}, they are indeed the same in the present case. 

For $g<\gamma$, $\braket{a^{\dagger}a}$ can be calculated exactly using either the Heisenberg equation or Keldysh formalism, as shown in Eq.\,\eqref{nonInteractingN}. Around the critical point ($g\lesssim \gamma$), we get the critical scaling for occupation number, 
\begin{equation}\label{eq:ncritscale}
\braket{a^{\dagger}a} \approx \frac{\gamma}{4} \frac{1}{\gamma-g},    
\end{equation}
which gives $\nu_{n}=1$.

Here we also show how to calculate $\nu_{n}$ with the quantum Langevin equation, as a demonstration of its application. From Eq.\,\eqref{thermalPx}, we know
\begin{equation}
\braket{x_{c}^2} = \frac{\int d {x_{c}} P(x_{c}) x_{c}^2 }{\int d {x_{c}} P(x_{c})} = \frac{\gamma}{\gamma-g}.
\end{equation}
Similarly, $\braket{p_{c}^2} = \gamma/(\gamma+g)$. Notice that $\braket{p_{c}^2}/\braket{x_{c}^2} = (\gamma-g)/(\gamma+g) \rightarrow 0$ as $g \rightarrow \gamma$, which shows the perfect squeezing of quadrature $p_{c}$. Then using $\braket{a^{\dagger}a} = \left( \braket{x_{c}^2}+\braket{p_{c}^2}-2 \right)/4$, we get again \eqref{eq:ncritscale} and therefore the same critical exponent $\nu_{n}=1$. 

\subsection{Dynamic critical exponent}\label{ScalingGap}

As seen in section \ref{sec:SpectrumL}, the gap $\Delta$ vanishes as the critical point is reached. Let $\Delta \sim (\gamma - g)^{\nu_{t}}$, where $\nu_{t}$ is called dynamic exponent since $\Delta$ is proportional to the decay rate of the correlation functions.

From Eq.\,\eqref{QLangevinEq}, we know that
\begin{equation}
 x_{c}(\omega)   =\frac{-4f_{x}(\omega)}{(\gamma-g - 2i\omega)},    
\end{equation}
which gives the correlation
\begin{equation}
\braket{x_{c}(t+\tau) x_{c}(t)} = \frac{\gamma}{\gamma-g} e^{-\frac{1}{2}(\gamma-g)|\tau|}. 
\end{equation}
It then follows that $\Delta \sim (\gamma-g) $ and so $\nu_{t}=1$.

The same result can also be obtained from the retarded Green function $g_{\text{R}}(\tau)$ \eqref{gRtau}, which also contains a vanishing decay rate $(\gamma-g)/2$ for large $\tau$.


\subsection{Finite-size scaling of occupation number}\label{FiniteSizeN}

For $g \lesssim \gamma$, the Kerr interaction can be ignored in the thermodynamic limit. However, to study the finite-size effects at the critical point $g=\gamma$, the Kerr interaction needs to be included. Here we show how to calculate finite-size effects based on the quantum Langevin equation.

From the saddle point solution (see Appendix \ref{Appdix:SaddlePoint}), at the critical point, $U|\alpha_{c}|^2$ is significant while $U|\alpha_{q}|^2$ vanishes. Then by ignoring irrelevant terms $\sim U|\alpha_{q}|^2$, the saddle point solution \eqref{saddle2} gives the Langevin equations
\begin{equation}\label{QLangevinU}
\begin{cases} \frac{d}{dt} x_{c} -\frac{U}{4} (x_{c}^2+p_{c}^2)p_{c} + \frac{1}{2} (\gamma-g) x_{c} + 2 f_{x} = 0,  \\[0.1cm]
\frac{d}{dt} p_{c} +\frac{U}{4} (x_{c}^2+p_{c}^2)x_{c} + \frac{1}{2} (\gamma+g) p_{c} + 2 f_{p} = 0.
\end{cases}
\end{equation}
Since $\braket{p_{c}^2}/\braket{x_{c}^2} \ll 1$ near the critical point, \eqref{QLangevinU} can be simplified to 
\begin{equation}\label{QLangevinUc}
\begin{cases} \frac{d}{dt} x_{c} -\frac{U}{4} x_{c}^2p_{c} + \frac{1}{2} (\gamma-g) x_{c}  + 2 f_{x} = 0,  \\[0.1cm]
\frac{d}{dt} p_{c} +\frac{U}{4} x_{c}^2x_{c} + \frac{1}{2} (\gamma+g) p_{c} + 2 f_{p} = 0,
\end{cases}
\end{equation}
where $(x_{c}^2+p_{c}^2) \approx x_{c}^2$ has been used. To get finite-size scaling of $\braket{a^{\dagger}a} \approx \braket{x_{c}^2}/4$, only the Langevin equation for $x_{c}$ needs to be considered. 

Setting $d p_{c}/dt = 0$ yields 
\begin{equation}
p_{c} = -\frac{U}{2(\gamma+g)} x_{c}^3 - \frac{4}{\gamma+g} f_{p}.
\end{equation}
Putting it back in \eqref{QLangevinUc}, we get 
\begin{equation}\label{QLangevinUcXc}
\frac{d}{dt} x_{c} + \frac{U^2}{8(\gamma+g)} x_{c}^5+ \frac{1}{2} (\gamma-g) x_{c}+ 2 f_{x} = 0,    
\end{equation} 
where $Ux_{c}^2 f_{p}$ has been ignored, since $Ux_{c}^2 \ll 1$ makes it negligible compared with the noise $2f_{x}$. Therefore, the finite-size effect introduces a weak sixth-order potential into the free energy ($|g-\gamma| \ll \gamma$):
\begin{equation}\label{FxU}
F(x_{c}) =  \frac{1}{4} (\gamma-g) x_{c}^2+ \frac{U^2}{48(\gamma+g)} x_{c}^6,
\end{equation}
which gives a clear analogue of Landau theory of symmetry breaking equilibrium phase transition. The unusually flat  $x_{c}^{6}$ potential originates from the coupling between $x_{c}$ and $p_{c}$ in \eqref{QLangevinUc}. Note that although the variable $x_{c}$ can be described using an effective thermal distribution, the system is intrinsically non-equilibrium as seen in the violation of the fluctuation-dissipation relation (see Appendix \ref{Appdix:FDR}).  At the critical point $g=\gamma$, using \eqref{thermalPx}, we find that 
\begin{equation}
\braket{x_{c}^2} = \frac{\sqrt{2\pi} }{\Gamma(7/6)} 3^{-2/3} \left( \frac{U}{\gamma}\right)^{-2/3},
\end{equation}
where $\Gamma(7/6) $ is the Gamma function. We then have the finite size scaling $\braket{a^{\dagger}a} \sim (U/\gamma)^{-\eta_{n}}$ with $\eta_{n}=2/3$.

\subsection{Finite-size scaling of gap}\label{FiniteSizeGap}

To get the finite-size scaling of the gap $\Delta$, we need to know the decay rate of the correlations. In the 
large $\tau$ limit ($\tau>0$), let
\begin{equation}
\braket{x_{c}(t+\tau)x_{c}(t)} \propto e^{-\kappa \tau},
\end{equation}
where $\kappa$ is an unknown constant that needs to be determined. Putting it into Eq.\,\eqref{QLangevinUcXc} at critical point, we get
\begin{equation}
\begin{split}
\kappa \braket{x_{c}(t+\tau)x_{c}(t)} = &\frac{U^2}{16\gamma} \braket{x_{c}^{5}(t+\tau)x_{c}(t)} \\&+ 2 \braket{f_{x}(t+\tau)x_{c}(t)}.
\end{split}
\end{equation}
We know $\braket{f_{x}(t+\tau)x_{c}(t)}=0$ due to causality. Since the free energy \eqref{FxU} has a weakly non-Gaussian potential, we can write
\begin{equation}
\begin{split}
\braket{x_{c}^{5}(t+\tau)x_{c}(t)} \approx\, & 15 \braket{x_{c}^{2}(t+\tau)}^2 \braket{x_{c}(t+\tau)x_{c}(t)} \\
&+ O(U^2/\gamma^2),
\end{split}
\end{equation}
where the leading order comes from Wick's theorem, while the higher-order corrections come from modification of Wick's theorem due to the weakly non-Gaussian term. Thus we find 
\begin{equation}
\kappa = \frac{15 U^2}{16\gamma} \braket{x_{c}^{2}}^2 = \gamma  \frac{5}{8} \frac{3^{-1/3} \pi}{\Gamma^{2}(7/6)}  \left( \frac{U}{\gamma}\right)^{2/3},
\end{equation}
which gives the finite-size scaling of gap  $\Delta/\gamma \propto \kappa/\gamma \sim (U/\gamma)^{\eta_{t}}$ with $\eta_{t}=2/3$.


\subsection{Argument using pseudo-critical point}
Finally, we want to show that the values of the critical exponents $\nu_{n}$, $\nu_{t}$, $\eta_{n}$, and $\eta_{t}$ are consistent with the argument using pseudo-critical point \cite{CardyFiniteSize2012}. Therefore, in practice, one can be obtained from the other three. 

With finite-size effects, we can define the pseudo-critical point $g_{c}(U)$ such that it replaces the real critical point $g_{c}(U=0)=\gamma$ in the scaling relations. Then, using scalings of $n$, we can have $n \sim \left|g_{c}(U)-g\right|^{-\nu_{n}}$ and $n \sim (U/\gamma)^{-\eta_{n}}$, which lead to the scaling 
\begin{equation}
\left|g_{c}(U)-g\right| \sim (U/\gamma)^{\eta_{n}/\nu_{n}}.   
\end{equation}
Further, using $\Delta \sim \left|g_{c}(U)-g\right|^{\nu_{t}} $, we get the finite-size scaling
\begin{equation}
\Delta  \sim (U/\gamma)^{\nu_{t}\eta_{n}/\nu_{n}}.   
\end{equation}
That is, the four exponents satisfy the relation
\begin{equation}
\eta_{t} \nu_{n}  =  \nu_{t} \eta_{n}, 
\end{equation}
which is indeed consistent with the above calculated exponents.

\section{Conclusion and Outlook}

Using a minimal model composed of a Kerr oscillator, two-photon driving, and single-photon loss, we study a second-order driven-dissipative quantum phase transition. We show that the phase transition is connected to the underlying $\mathcal{PT}$ symmetry of the semi-classical dynamics, which provides a stabilization mechanism. We anticipate that there exists a class of phase transitions with a similar underlying mechanism. Quantum fluctuations are studied using ED and the Keldysh formalism. Critical properties such as symmetry breaking, finite-size scaling, state squeezing, and spectral properties are explored. We show that the emergence of criticality in this system is due to perfect squeezing at the critical parametric driving. Critical scaling and finite-size scaling properties are calculated analytically using the quantum Langevin equation. Since an analytical solution is provided, this system can serve as a paradigmatic platform for the exploration of open quantum many-body physics.

\begin{acknowledgments}
We thank Thomas Barthel, Hao Geng, and Jian-Guo Liu for helpful conversations, and Wouter Verstraelen for valuable comments on the manuscript. We thank the anonymous referee B for bringing the finite-size analysis in Ref.\,\cite{TorrePRA2013} to our attention. This work was supported by the U.S.\,Department of Energy, Office of Science, Office of Basic Energy Sciences, Materials Sciences and Engineering Division under Award No.\,DE-SC0005237. XHHZ was also supported in part by Brookhaven National Laboratory's LDRD project No.\,20-024.
\end{acknowledgments}

\appendix


\section{The vectorization of the density matrix and exact diagonalization (ED)}\label{Appdix:ED}

Using an auxiliary Hilbert space of the same dimension as that of the original one $\mathcal{H}$, the density matrix $\rho = \sum_{ij} \rho_{ij} \ket{i}\bra{j}$ can be written as a vector $\vec{\rho} = \sum_{ij} \rho_{ij} \ket{i}\otimes \ket{j}$ in the enlarged space $\mathcal{H} \otimes \mathcal{H}$. Then the Lindblad superoperator 
\begin{equation}
\mathcal{L} (\bullet) = -i [H,\bullet] + \sum_{k} \gamma_{k} \Big( c_{k} \bullet c_{k}^{\dagger} - \frac{1}{2} \{ \bullet, c_{k}^{\dagger} c_{k} \} \Big)
\end{equation}
can be written as an operator $L$ in $\mathcal{H} \otimes \mathcal{H}$, whose form is given by
\begin{equation}
\begin{split}
\hat{L} = &-i H\otimes \mathbb{1} + i \mathbb{1} \otimes H^{T} \\ &+ \sum_{k} \gamma_{k} \Big( c_{k} \otimes c_{k}^{*} - \frac{1}{2} \mathbb{1} \otimes  c_{k}^{T}c_{k}^{*} - \frac{1}{2}  c_{k}^{\dagger} c_{k} \otimes \mathbb{1}  \Big).
\end{split}
\end{equation}

In our case, the jump operator $c_{k}$ is a boson annihilation operator $a$. In numerical calculation, a cutoff for the maximum occupation number $n_{\text{max}}$ is chosen such that $a^{\dagger} \ket{n_{\text{max}}} =0$. Since the occupation number is known to be of order $1/U$ in our system, $n_{\text{max}}$ is chosen be of the same order and increased gradually until cutoff errors are acceptable.

ED of $\hat{L}$ gives eigenstates $\{ \sigma_{i} \}$ and their corresponding eigenvalues. Some of the eigenstates are not positive semidefinite, which means they cannot be physical density matrices. For the physical states, $\sigma$ is normalized by their trace i.e.\,$\sigma \rightarrow 
\sigma/\Tr (\sigma)$. For the non-physical states such as the $\sigma_{1}$ in Fig.\,\ref{Wigner} of the main text, we use the trace norm $\Tr(|\sigma|)$,  i.e.\,$\sigma \rightarrow 
\sigma/\Tr (|\sigma|)$.

\section{Keldysh action in the original basis}\label{Appdix:Keldysh}

By writing the Lindbladian as path integral using a coherent state basis \cite{SiebererRPP2016}, the partition function for our model can be written as
\begin{equation}
Z = \Tr[\rho(+\infty)]  = \int D[\alpha_{+}(t),\alpha_{+}^{\star}(t),\alpha_{-}(t),\alpha_{-}^{\star}(t)] \,\, e^{i S},
\end{equation}
where $\alpha_{+}$ and $\alpha_{-}$ are the basis in the ket and bra branches respectively. The action $S$ is given by
\begin{equation} 
\begin{split}
S =& \int dt \, \big[ \alpha_{+}^{*} i \frac{\partial }{\partial t} \alpha_{+} - \alpha_{-}^{*}  i \frac{\partial }{\partial t} \alpha_{-} - (H_{+} - H_{-}) \\ &-i \gamma \alpha_{+} \alpha_{-}^{*} +i \frac{\gamma}{2} \alpha_{+}^{*} \alpha_{+} +i \frac{\gamma}{2} \alpha_{-}^{*} \alpha_{-} \big],
\end{split}
\end{equation} 
where $H_{\pm}$ are obtained by replacing operators $a$ and $a^{\dagger}$ in $H$ by $\alpha_{+/-}$ and $\alpha_{+/-}^{*}$. The action in Eq.\,\eqref{Scq} can be then obtained with a Keldysh rotation to the basis $\alpha_{c/q}$.

\section{Saddle point solution} \label{Appdix:SaddlePoint}

To get the saddle point solution, we apply functional derivatives to get
\begin{widetext}
\begin{eqnarray}
\frac{\delta }{\delta \alpha_{c}^{*}} S &=& i \frac{\partial }{\partial t} \alpha_{q} - \frac{U}{2} \alpha_{c}^2 \alpha_{q}^{*} - U |\alpha_{c}|^2 \alpha_{q} - \frac{U}{2} |\alpha_{q}|^2 \alpha_{q} - \frac{G}{2} \alpha_{q}^{*} - i \frac{\gamma}{2} \alpha_{q} =0 \label{saddle1}\\
\frac{\delta }{\delta \alpha_{q}^{*}} S &=& i \frac{\partial }{\partial t} \alpha_{c} - \frac{U}{2} |\alpha_{c}|^2 \alpha_{c} - U |\alpha_{q}|^2 \alpha_{c} - \frac{U}{2} \alpha_{q}^2 \alpha_{c}^{*} - \frac{G}{2} \alpha_{c}^{*} + i \frac{\gamma}{2} \alpha_{q} + i \frac{\gamma}{2} \alpha_{c}  =0 \label{saddle2}\\
\frac{\delta }{\delta \alpha_{c}}     S &=& -i \frac{\partial }{\partial t} \alpha_{q}^{*} - \frac{U}{2} (\alpha_{c}^{*})^2 \alpha_{q}^{*} - U |\alpha_{c}|^2 \alpha_{q}^{*} - \frac{U}{2} |\alpha_{q}|^2 \alpha_{q}^{*} - \frac{G^{*}}{2} \alpha_{q} + i \frac{\gamma}{2} \alpha_{q}^{*} =0 \label{saddle3}\\
\frac{\delta }{\delta \alpha_{q}}     S &=& -i \frac{\partial }{\partial t} \alpha_{c}^{*} - \frac{U}{2} |\alpha_{c}|^2 \alpha_{c}^{*} - U |\alpha_{q}|^2 \alpha_{c}^{*} - \frac{U}{2} (\alpha_{q}^{*})^2 \alpha_{c} - \frac{G^{*}}{2} \alpha_{c} + i \frac{\gamma}{2} \alpha_{q}^{*} - i \frac{\gamma}{2} \alpha_{c}^{*}  =0 \label{saddle4}
\end{eqnarray}
\end{widetext}

From the requirements $ \eqref{saddle1} =  \eqref{saddle3}^{*} $ and  $ \eqref{saddle2} =  \eqref{saddle4}^{*} $, we get $\bar{\alpha}_{q}=0$ and
\begin{equation}\label{saddle}
\frac{\partial }{\partial t} \bar{\alpha}_{c} = -i \frac{U}{2} |\bar{\alpha}_{c}|^2 \bar{\alpha}_{c}- i \frac{G}{2} \bar{\alpha}_{c}^{*} - \frac{\gamma}{2} \bar{\alpha}_{c}, 
\end{equation}
which is exactly the mean-field equation of motion by identifying $\alpha_{\text{MF}} = \alpha_{c} / \sqrt{2} = (\alpha_{+} + \alpha_{-})/2$. 

\section{non-Gaussian part} \label{Appdix:nonGaussian}

By expanding around the saddle point solution, we obtain the action of fluctuations $S=S_{\text{G}}+S_{\text{NG}}$, where $S_{\text{G}}$ is the Gaussian part shown in Eq.\,\eqref{SG} of the main text and $S_{\text{NG}}$  is the non-Gaussian part: 
\begin{equation}
\begin{split}
S_{\text{NG}} =& \int dt\,  (-\frac{U}{2}) \Big[  |\delta\alpha_{c}|^2 ( \bar{\alpha}_{c}^{*}  \delta\alpha_{q} + \text{c.c.}  ) 
\\ &+ \big( \bar{\alpha}_{c} (\delta\alpha_{c}^{*})^2  \delta\alpha_{q} + \text{c.c.} \big)  \\
&+ ( |\delta\alpha_{c}|^2 + |\delta\alpha_{q}|^2  ) ( \bar{\alpha}_{c} \delta\alpha_{q}^{*} + \text{c.c.})    \\
&+( |\delta\alpha_{c}|^2 + |\delta\alpha_{q}|^2  ) (\delta\alpha_{c} \delta\alpha_{q}^{*} +\text{c.c.}  )  \Big].
\end{split}
\end{equation}
The non-quadratic fluctuations are of the form $U\alpha^{*}_{c} \delta\alpha^{*}_{\mu_1} \delta\alpha_{\mu_2} \delta\alpha_{\mu_3}  $ and $U \delta\alpha^{*}_{\mu_1} \delta\alpha^{*}_{\mu_2} \delta\alpha_{\mu_3} \delta\alpha_{\mu_4}$. In the thermodynamic limit $U\rightarrow 0^{+}$, since $\bar{\alpha}_{c} \sim 1/\sqrt{U}$, the third-order fluctuations are $\sim \sqrt{U}$ and the fourth-order fluctuations are $\sim U$ while the Gaussian fluctuations are $\sim 1$. Therefore, $S_{\text{NG}}$ can be neglected in the thermodynamic limit.


\section{The calculation of the Green functions} \label{Appdix:GreenFunction}

The Green functions we consider are the Keldysh, advanced, and retarded Green function: 
\begin{equation}
\begin{split}
g_{\text{K}}(t_1,t_2) &\equiv -i \braket{\alpha^{*}_{c} (t_1) \alpha_{c}(t_2) } \\ &= -i\braket{ \{a^{\dagger}(t_1), a(t_2) \} },\\
g_{\text{A}}(t_1,t_2) &\equiv -i \braket{\alpha^{*}_{c} (t_1) \alpha_{q}(t_2) } \\ &= -i \theta(- t_2 + t_1 ) \braket{ [ a^{\dagger}(t_1), a(t_2) ] }, \\
g_{\text{R}}(t_1,t_2) &\equiv -i \braket{\alpha^{*}_{q} (t_1) \alpha_{c}(t_2) } \\&= -i \theta(t_2 - t_1) \braket{ [ a(t_2), a^{\dagger}(t_1) ] },
\end{split}
\end{equation}
while $\braket{\alpha^{*}_{q} (t_1) \alpha_{q}(t_2) } = 0$. In the steady state, they only depend on the time difference $\tau=t_2-t_1$. 

For a Gaussian theory like Eq.\,\eqref{SG}, we can obtain Green functions analytically \cite{AltlandSimonsBook2010}. Using the Fourier transform $\alpha(\omega) = \frac{1}{\sqrt{2\pi}} \int dt \, \alpha(t) e^{i \omega t} $, Eq.\,\eqref{SG} can be written in frequency space as 
\begin{equation}\label{SGmatrix}
S_{G} = \int d\omega \Psi^{\dagger}(\omega) \frac{i}{2} A (\omega) \Psi(\omega),
\end{equation}
where 
\begin{equation}
\Psi(\omega) \equiv [ \delta\alpha_{c}(\omega) ,\, \delta\alpha_{c}^{*}(-\omega) ,\, \delta\alpha_{q}(\omega) ,\, \delta\alpha_{q}^{*}(-\omega)]^{T},
\end{equation} 
and
\begin{widetext}
\begin{equation}
A(\omega)=-i 
\begin{bmatrix}
0 & 0 & \omega - i \frac{\gamma}{2} - U |\bar{\alpha}_{c}|^2 & -\frac{G}{2} - \frac{U}{2} \bar{\alpha}_{c}^2 \\
0 & 0 & -\frac{G^{*}}{2}- \frac{U}{2} (\bar{\alpha}_{c}^{*})^2 & -\omega+i \frac{\gamma}{2} -U |\bar{\alpha}_{c}|^2   \\
\omega + i \frac{\gamma}{2} - U |\bar{\alpha}_{c}|^2 & -\frac{G}{2} - \frac{U}{2} \bar{\alpha}_{c}^2 & i \gamma & 0  \\
-\frac{G^{*}}{2} - \frac{U}{2} (\bar{\alpha}_{c}^{*})^2 & -\omega - i \frac{\gamma}{2} - U |\bar{\alpha}_{c}|^2 & 0 & i \gamma 
\end{bmatrix} 
\equiv
\begin{bmatrix}
0 & P^{\text{A}}   \\
P^{\text{R}} & P^{\text{K}}
\end{bmatrix}.
\end{equation}
\end{widetext}
Note here that the four components of $\Psi(\omega)$ are four independent variables. The action below the critical point can be obtained by setting $\bar{\alpha}_{c}=0$.

Define $C=(A^{-1})^{T}$ and
\begin{equation}
C
\equiv
\begin{bmatrix}
C^{\text{K}} & C^{\text{R}}   \\
C^{\text{A}} & 0
\end{bmatrix}.
\end{equation}
We can see that $C^{A} = \big[(P^{A})^{-1}\big]^{T}$, $C^{R} = \big[(P^{R})^{-1}\big]^{T} $, and $C^{K} = \big[ - (P^{R})^{-1} P^{K} (P^{A})^{-1} \big]^{T}$. Then the three Greens functions needed here can be obtained by identifying
\begin{eqnarray}
g_{\text{K}/\text{A}/\text{R}} (\omega) &=& -i \big(C^{\text{K}/\text{A}/\text{R}}(\omega)\big)_{11},
\end{eqnarray}
where $g_{\text{R}}^{*} (\omega) = g_{\text{A}} (\omega)$.
For example, above critical point,
\begin{equation}
i g_{\text{K}} (\omega) = \frac{\gamma}{4}\frac{4\omega^2 + 16 \phi \omega + 16 \phi^2 + 2 \gamma^2}{ \omega^4 + (\gamma^2-8\phi^2)\omega^2 + 16 \phi^4}
\end{equation}
and
\begin{equation}
i g_{\text{R}} (\omega) = \frac{2 \left( \gamma - 2i \omega -4i \phi  \right)}{(\gamma - 2i \omega)^2+ 16 \phi^2 - \gamma^2 }.
\end{equation}

In real space,
\begin{equation}\label{GreenFunctionTau}
g_{\text{K}/\text{A}/\text{R}} (\tau) = \frac{1}{2\pi} \int d\omega\, g_{\text{K}/\text{A}/\text{R}} (\omega) e^{-i \omega \tau}. 
\end{equation}


\section{Critical exponents from above}\label{Appdix:ExponentAbove}

Using the Green function \eqref{GreenFunctionTau}, we get that, for $g \gtrsim \gamma$, 
\begin{equation}
i g_{\text{K}} (\tau) \approx \frac{\gamma}{8(g-\gamma)} e^{-2(g-\gamma)|\tau|} + \frac{1}{4} e^{-\gamma|\tau|} , 
\end{equation}
where expansion to lowest order of $(g-\gamma)$ has been applied. It can be seen that $\Delta \sim (g-\gamma)^{\nu_{t}^{\prime}}$, where $\nu_{t}^{\prime}=1 =\nu_{t}$. Since $n \sim ig_{\text{K}}(\tau=0) \sim (g-\gamma)^{-\nu_{n}^{\prime}}$, we have $\nu_{n}^{\prime}=1=\nu_{n}$.

\section{Violation of the Fluctuation-Dissipation Relation}\label{Appdix:FDR}

Although $x_{c}$ itself can be described by an effective thermal equilibrium distribution, the system is intrinsically non-equilibrium when considering all variables $x_{c}$, $p_{c}$, $x_{q}$, and $p_{q}$ (e.g.\,\cite{TorrePRA2013,SiebererRPP2016}). The non-equilibrium nature can be shown in the violation of the fluctuation-dissipation relation. In thermal equilibrium at temperature $T$, we have the fluctuation-dissipation relation (e.g.\,\cite{SiebererRPP2016,KamenevBook2011,StefanuccivanLeeuwen2013}) 
\begin{equation}
    g_{\text{K}} (\omega) = h(\omega) \left( g_{\text{R}} (\omega) - g_{\text{A}} (\omega) \right),
\end{equation}
where $h(\omega) = \coth \frac{\omega}{2T} = 2 n_{\text{B}} (\omega/T) +1$ with the Bose-Einstein distribution $n_{\text{B}} (\omega/T) = 1/(e^{\omega/T}-1)$. In our case, we can define
\begin{equation}
\begin{split}
    \tilde{h}(\omega) &= \frac{ g_{\text{K}} (\omega) }{ g_{\text{R}} (\omega) - g_{\text{A}}(\omega) } = \frac{1}{\pi \gamma} \frac{S_{\text{inel}}(\omega)}{\mathcal{A}(\omega)} + 1. 
\end{split}
\end{equation}
Then we know
\begin{equation}
\begin{split}
    \tilde{h}(\omega) =   \begin{cases} 
\dfrac{2|G|^2}{ 4\omega^2 + \gamma^2-|G|^2 }+1  & |G|<\gamma, \\[0.3cm]
\dfrac{\gamma^2}{2} \dfrac{1}{(\omega+2\phi)^2}+1 & |G|>\gamma .
\end{cases} 
\end{split}
\end{equation}
$\tilde{h}(\omega)$ cannot be given by a thermal distribution when $|G|\neq 0$. Without driving, the system is indeed in thermal equilibrium with temperature $T=0$.   


\bibliography{DPT}

\end{document}